\documentclass[draft,jgrga]{agutex}
  \usepackage{graphicx}
 \setkeys{Gin}{draft=false}
\authorrunninghead{HU ET AL.}

\titlerunninghead{MAGNETIC FIELD-LINE LENGTHS IN FLUX ROPES}

\begin{document}

%% ------------------------------------------------------------------------ %%
%
%  TITLE
%
%% ------------------------------------------------------------------------ %%

\title{Magnetic field-line lengths inside interplanetary magnetic flux ropes}
%
% e.g., \title{Terrestrial ring current:
% Origin, formation, and decay $\alpha\beta\Gamma\Delta$}
%

%% ------------------------------------------------------------------------ %%
%
%  AUTHORS AND AFFILIATIONS
%
%% ------------------------------------------------------------------------ %%

%Use \author{\altaffilmark{}} and \altaffiltext{}

% \altaffilmark will produce footnote;
% matching \altaffiltext will appear at bottom of page.

 \authors{Qiang Hu,\altaffilmark{1}
 Jiong Qiu,\altaffilmark{2} and Sam Krucker\altaffilmark{3}}
% John B. McDougall\altaffilmark{4}, and S. Visconti\altaffilmark{5}}

\altaffiltext{1}{Department of Space Science and CSPAR, The
University of Alabama in Huntsville,  Huntsville, AL, USA.}

\altaffiltext{2}{Department of Physics, Montana State University,
Bozeman, MT, USA.}

\altaffiltext{3}{Space Science Laboratory, University of
California, Berkeley, CA, USA.}

%\altaffiltext{4}{Division of Hydrologic Sciences, Desert Research
%Institute, Reno, Nevada, USA.}

%\altaffiltext{5}{Dipartimento di Idraulica, Trasporti ed
%Infrastrutture Civili, Politecnico di Torino, Turin, Italy.}

%% ------------------------------------------------------------------------ %%
%
%  ABSTRACT
%
%% ------------------------------------------------------------------------ %%

% >> Do NOT include any \begin...\end commands within
% >> the body of the abstract.

\begin{abstract}
We report on the detailed and systematic study of field-line twist
and length distributions within magnetic flux ropes embedded in
Interplanetary Coronal Mass Ejections (ICMEs). The Grad-Shafranov
reconstruction method is utilized together with a constant-twist
nonlinear force-free (Gold-Hoyle) flux rope model to reveal the close relation between the
field-line twist and length in cylindrical flux ropes, based on
in-situ Wind spacecraft measurements. We show that the field-line
twist distributions within interplanetary flux ropes are
inconsistent with the Lundquist model. In particular we utilize
the unique measurements of magnetic field-line lengths within
selected ICME events as provided by \citet{2011JGRAK} based on
energetic electron burst observations at 1 AU and the associated
type III radio emissions detected by the Wind spacecraft. These
direct measurements are compared with our model calculations to
help assess the flux-rope interpretation of the embedded magnetic
structures. By using the different flux-rope models, we show that
the in-situ direct measurements of field-line lengths are
consistent with a flux-rope structure with spiral field lines of
constant and low twist, largely different from that of the
Lundquist model, especially for relatively large-scale flux ropes.
%We therefore provide support to advocate the constant-twist
%flux-rope model for fitting in-situ spacecraft observations of
%magnetic flux ropes.
\end{abstract}

%% ------------------------------------------------------------------------ %%
%
%  BEGIN ARTICLE
%
%% ------------------------------------------------------------------------ %%

% The body of the article must start with a \begin{article} command
%
% \end{article} must follow the references section, before the figures
%  and tables.

\begin{article}

%% ------------------------------------------------------------------------ %%
%
%  TEXT
%
%% ------------------------------------------------------------------------ %%

\section{Introduction}\label{sec:intro}
Magnetic flux ropes are a type of well organized magnetic field
structures embedded in space plasmas. The existence of such
structures is best confirmed by in-situ spacecraft observations
and the associated modeling when the structure is traversed by the
spacecraft \citep[e.g.,][]{1995ISAAB,Lepping1990,Lepping1997}. In
addition, many studies on the origination of such structures also
provide mostly indirect evidence to support such interpretation of
these structures as magnetic flux ropes
\citep[e.g.,][]{Webb2000,Longcope2007b,Qiu2007,2008AnGeoD,Qiu2009,Vourlidas2014}.
They are found in Interplanetary Coronal Mass Ejections (ICMEs),
the interplanetary counterparts of CMEs originating from the Sun.

Some ICMEs are traditionally categorized as Magnetic Clouds (MCs)
that possess a specific set of signatures based on in-situ
spacecraft measurements of both magnetic field and bulk plasma
properties. A more modern view of all ICMEs containing flux ropes
is also emerging \citep{2013SoPh1G,2013SoPhG,2013SoPhX,2014ApJH}.
This seems reasonable especially considering that most origination
mechanisms for CMEs involve magnetic flux ropes no matter whether
they are considered to be pre-existing prior to eruption, or
generated during the process. Moreover the subsequent argument is
that such structures originating from the Sun and propagating into
the interplanetary space may not be properly detected by the
in-situ spacecraft. Each spacecraft only provides a very
localized, single-point measurements of the structure traversed.
Therefore depending on the relative spacecraft path across the
structure, the variability and limitation in the in-situ
signatures of magnetic flux ropes are significant, resulting in
the incidences when the flux-rope structure is present, but the
in-situ signatures are lacking \citep[e.g.,][]{2006SoPhJL}.
However if one adheres to the traditional definitions of MCs,
which satisfies these criteria: 1) relatively strong magnetic
field magnitude, 2) smooth rotation in magnetic field direction,
and 3) relatively low proton $\beta$, the ratio between the plasma
pressure and the magnetic pressure, one can likely derive a
magnetic flux-rope structure from the in-situ data.

Effort has been put on in-situ modeling of magnetic flux-rope
structures in order to extend the current capability thus to
better reveal and characterize these structures in a quantitative
manner. Various flux-rope models utilize in-situ spacecraft
measurements of magnetic field and plasma parameters along the
spacecraft path and are based on either a cylindrical or toroidal
geometry and magnetohydrostatic theory. They range from the
well-known one-dimensional (1D) linear force-free cylindrical
model \citep{lund}, to the corresponding toroidal model
\citep{2007AnGeoM,2003GeoRLR}, and to the fully two and a half dimensional
($2\frac{1}{2}$D) Grad-Shafranov (GS)  reconstruction model
\citep{2002JGRAHu}. One particular model that has not been widely
recognized is the so-called Gold-Hoyle (GH) model that was
originally developed by \citet{1960MNRAS.GH} and was only applied
in a limited number of studies \citep{1999AIPCF,Dasso2006,2014ApJH}.
The distinct features of this model, remaining 1D, are that the
field-line twist is constant across the radius and the
corresponding equilibrium state is non-linear force free.

In our latest study of \citet{2014ApJH}, we showed that the
flux-rope structures as derived from the generally non-force free
GS method are more consistent with the GH model than with the
Lundquist model, especially in that the field-line twist
distributions within ICME flux ropes remain fairly constant for
large-size, low-twist flux ropes. In the present study, we intend
to elaborate more on this finding and present additional
consistency check by utilizing the unique measurements of
field-line lengths inside MCs.

A unique set of in-situ spacecraft observations besides the
magnetic field and plasma parameters in interplanetary space is
the energetic electron burst onset. They appear as sudden increase
in electron flux of energies up to a few hundred keV \citep{1999ApJK,2006ApJK,2011ApJW} as
the electron beams propagate away from the source on the Sun to the
location of the spacecraft along individual field lines connecting
both ends. Under certain assumptions such as scatter-free
propagation and coincidental release at the time of associated
Type III radio burst, the path lengths of magnetic field lines can
be derived especially inside MCs. There are two ways to obtain the
length estimate based on electron burst onset observations: one is
to directly calculate the length traveled by the product of the
speed of electrons (of known energy) and the travel time (taken as
the difference between the onset time at 1 AU and the release time
as given by the corresponding Type III onset time); the other is
to linearly fit the onset times  of electrons of different
energies versus their inverse speeds (so-called inverse-beta
method; \citep{2006ApJK}) and the slope yields the path length.  The first
study of comparing field-line path lengths inside an MC utilizing
the electron burst measurements was carried out by
\citet{1997GeoRLL}. They combined multiple in-situ measurements
from the Wind spacecraft during an MC interval to derive
field-line lengths as measured by the energetic electrons travel
time multiplied by the speed which were then compared with the
lengths estimated based on certain flux-rope models of MCs. They
found for one particular event that the path lengths at several
locations inside the MC ranging from about 1.2 AU near the center
to about 3 AU near the boundary, consistent with flux-rope model
estimates. Such type of study, rare but important, provided unique
and direct supporting evidence for the interpretation of MC
structures as magnetic flux ropes. Not until recently did
\citet{2011JGRAK,2011ApJK} extended that unique study by applying the same
analysis to a set of Wind MC events and additional electron burst events from the ACE spacecraft. They derived the field-line
lengths based on in-situ electron burst onset and associated Type
III radio burst following the approach of \citet{1997GeoRLL} and
compared with two flux-rope models: one being the Lundquist model
and the other flux-conservation model given in \citet{1997GeoRLL}.
Their comparison indicated poor correlation between the measured
and the model field-line lengths with the latter being exceedingly
larger, $\ge 4$ AU with maxima reaching about 10 AU, especially
for the Lundquist model. Their results cast doubt on the model fit
to MC flux ropes by the Lundquist model which intrinsically
possesses the property of increasing field-line twist thus length
from the center towards the boundary of the flux rope at a rapid
rate, approaching infinity at the boundary defined as a circular
cylindrical surface of vanishing axial magnetic field. In
addition, our own analysis \citep{2014ApJH} also showed that the
field-line twist estimates from the GS method are not consistent
with the Lundquist model but more aligned with the GH model of
constant twist. In the present study, we will focus our analysis
on the field-line length estimates based on the GS reconstruction
results, supplemented by the corresponding estimates based on the
GH model as well.

Estimates of magnetic field-line lengths, by taking advantage of the
unique and comprehensive in-situ spacecraft measurements, not only
provide constraint and validation of flux-rope models, but also
provide possible measurement of one key parameter, the axial
length of a cylindrical flux rope. This parameter determines the
quantitative measurements of the poloidal magnetic flux and the
relative magnetic helicity contents \citep{Qiu2007,2010JGRAW}. Since all existing flux-rope
models based on in-situ measurements are 2D at best in geometry,
significant uncertainty exists for the axial dimension. An
effective axial length, $L_{eff}$, has to be used in order to
determine the quantities of poloidal magnetic flux and relative magnetic
helicity within a cylinder of finite length $L_{eff}$ that are equivalent to the
corresponding content contained within the actual flux-rope structure. In
our effort to connect the ICME flux ropes with their solar source
region properties, particularly by comparing the magnetic flux
contents at both ends, a somewhat arbitrary range of $L_{eff}\in[0.5,2.0]$ AU was
used \citep{Qiu2007,2014ApJH}. We strive to gain more insight and to
obtain a refined range of effective axial length from the current analysis of magnetic field-line length estimates inside
MCs.

The article is organized as follows. We present the detailed
description of magnetic field-line length estimates inside the MCs
in the next section, for the Wind spacecraft MC events given by
\citet{2011JGRAK} for which the length measurements based on
electron burst onset were published. We will reconstruct the
structures of these events and derive the relevant characteristic parameters
by using the GS method and the GH model will be primarily utilized
to provide extrapolated estimates on field-line lengths. The
approach of obtaining various length estimates is described in
Section~\ref{sec:length}. These estimates are compared one-by-one
with the corresponding electron burst measurements from the Wind
spacecraft. Three cases are selected to be presented in detail in
Section~\ref{sec:CS}. A summary  of our results and comparison for
all events is given in Section~\ref{sec:summ}. We finally conclude
and discuss the implications of our results in the last section.

\section{Magnetic Field-line Length Estimates}\label{sec:length}
We re-examine the events listed in Table~1 of \citet{2011JGRAK},
total of 8 Wind spacecraft MC events with given electron burst
measurements. We are able to successfully reconstruct 7 MC events
by the GS method, except for the one on 2 May 1998. Therefore this
event is excluded from our analysis. In addition, only the
measurements of electron burst events occurring in the identified
GS intervals (see Table~\ref{tbl:GS}) are utilized in our
analysis. Others falling out of the GS intervals are excluded as
well.

\subsection{GS Reconstruction Results}\label{sec:GS}
The GS reconstruction is to solve the plane Grad-Shafranov (GS)
equation of the magnetic flux function $A(x,y)$ (equivalent to the
$z$ component of the magnetic vector potential; in direct analogy
to a stream function in two-dimensional incompressible flow) on
the cross-section of a cylindrical flux rope. Therefore the
transverse magnetic field $(B_x,B_y)$ is fully determined by the
flux function $A$ and the magnetic field lines are lying on
cylindrical iso-surfaces of $A$, called $A$ shells. The
non-vanishing axial magnetic field component $B_z$ becomes a
function of $A$ only, yielding a cylindrical but non-axisymmetric
flux-rope configuration. Detailed descriptions of the method and
the latest updates were reported in
\citet{2002JGRAHu,2013SoPhH,2014ApJH}.

\begin{figure}
 \noindent\includegraphics[width=8.3cm]{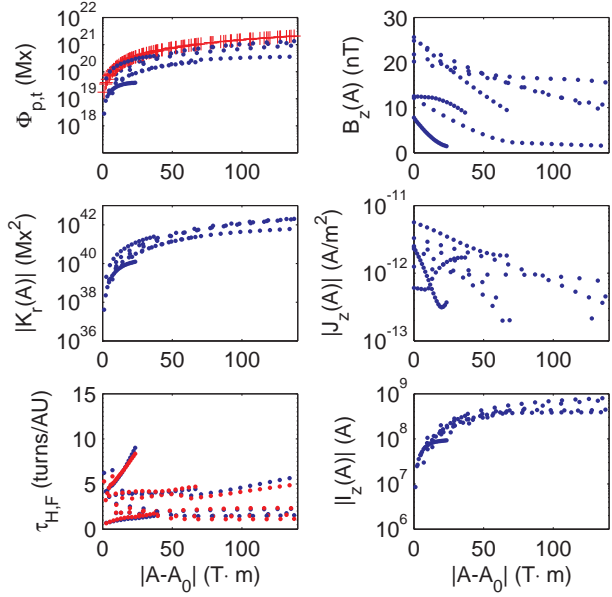}
 \caption{ Summary plot of various physical quantities (unsigned) vs. the shifted flux function for the Wind MC events:
 (counterclock-wise from the top left panel)
the poloidal (red pluses) and toroidal magnetic flux $\Phi_{p,t}$,
the  relative magnetic helicity, the field-line twist estimates
$\tau_H$ (red dots) and $\tau_F$ (blue dots) \citep{2014ApJH}, the axial current,
the axial current density, and the axial magnetic field.}
 \label{fig:var_e}
 \end{figure}

Various physical quantities characterizing such a flux-rope
structure can be derived including the axial magnetic field $B_z$,
the axial electric current density and current, the toroidal
(axial) and poloidal magnetic flux $\Phi_{t,p}$, the relative
magnetic helicity $K_r$, and the magnetic field-line twist. They
are all functions of $A$ alone and vary across distinct $A$
shells. Figure~\ref{fig:var_e} shows the summary plots of these
quantities as they vary along the $A$ shells for all the Wind MC events we
examined. They generally exhibit a similar pattern to the other
flux-rope events as we first reported in \citet{2014ApJH} in this
congregated manner. The magnetic fluxes increase monotonically
from the center toward the outer boundary while the poloidal flux
is generally larger than the toroidal flux. The relative magnetic
helicity also increases monotonically and smoothly. So does the
electric current since they are all accumulative integral
quantities. The axial field, on the other hand, shows a
monotonically declining profile from the center outwards, typical
of a flux-rope structure. The maximum value ($B_{z0}$) ranges
between a few and a few tens nT. The most irregular variation
exists in the current density which represents the first-order
derivative of a transverse pressure with respect to $A$.  The
field-line twist estimates displayed here, i.e.,
$\tau_H=|K_r|/\Phi_t^2$ and $\tau_F=\Phi_p/\Phi_t$,
are only for qualitative visual inspection since they are less
reliable as we discussed in \citet{2014ApJH}. Several scalar
quantities representing the total magnetic flux and magnetic
helicity contents within certain boundary $A=A_b$ are given in
Table~\ref{tbl:GS} together with the approximate average twist
estimates $\bar\tau_H,\bar\tau_F$. The other estimates for average twist ($\langle\tau\rangle$ and $\tau_0$) are based on more
quantitatively reliable calculations to be described below.

\begin{figure}
 \noindent\includegraphics[width=8.3cm]{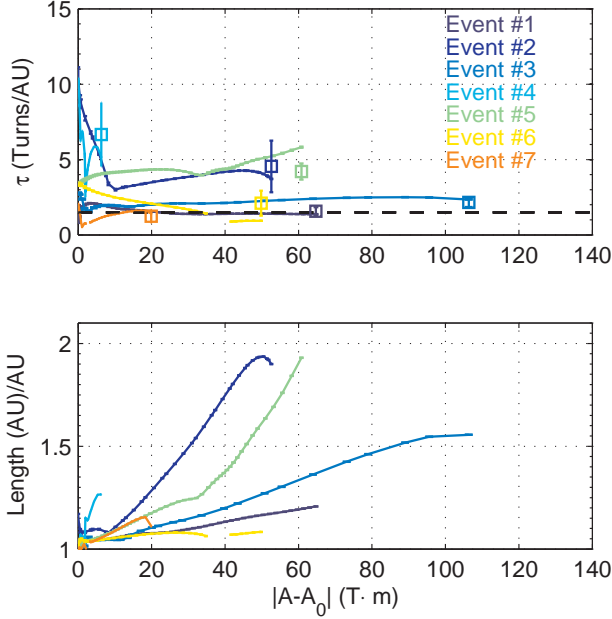}
 \caption{Summary plot of field-line twist $\tau$ (top) and length (bottom) distributions (both for $L_{eff}=1$ AU) vs. the shifted flux function for the Wind MC events. Different colors represent different events as indicated by
 the legend in the top panel. The square symbol and associated error bar at the end of each curve where $A'\equiv A_c$ indicate the mean and the standard deviation of each curve.}
 \label{fig:tau_e}
 \end{figure}

In our latest study of \citet{2014ApJH}, we performed systematic
study of field-line twist distribution within ICME flux ropes based
on the GS reconstruction method. The field-line twist, $\tau(A)$,
also as a single-variable function of $A$, is obtained by the
graphic method described in \citet{2014ApJH}. That is for each
individual field line lying on a distinct surface of one particular $A$
value, usually an open-ended cylindrical surface of closed side,
denote the axial length along which the field line completes one
full turn, $L_z$ in AU, then the field-line twist is simply
\begin{equation}
\tau(A)=\frac{1}{L_z(A)},
\end{equation}
in unit of turns/AU. The field-line length $L_s$ is easily obtained by
the line integral along each individual field line, i.e., by summing up the
distances between the adjacent end points of each line segment.
The results for the 7 events by this approach are shown in
Figure~\ref{fig:tau_e}. In the top panel, the mean value of each
curve (each event) $\langle\tau(A)\rangle$ and the corresponding
standard deviation $\Delta\tau$ as given in Table~\ref{tbl:GS} are
shown as the square symbol and the associated  error bar,
respectively. They  generally indicate that for most events, the twist
remains fairly constant, excluding the region close to the center
where the shifted flux function $A'=|A-A_0|=0$. Correspondingly, the length distribution shows a viable rate of increase with increasing $A'$.
  Note that each
curve ends at certain value of $A$ or  $A'=A_c$ corresponding to
the outermost closed loops on the flux-rope cross-section,
represented by the equi-value contours of $A$. Beyond this
boundary, the field lines can no longer complete one full turn
within the computational box. Hence no field-line twist and length
estimates by the graphic method are available \citep{2014ApJH}.

To circumvent this limitation and after observing that the twist
distributions exhibit a trend of remaining fairly  constant
throughout the outer region of a flux rope, as first reported in
\citet{2014ApJH} and further demonstrated here, we employ a
theoretical, constant-twist flux-rope model to assist in the
analysis. To reinforce and justify this additional approach, we
put the results for all the events we have examined in
\citet{2014ApJH} and the present study together onto
Figure~\ref{fig:tau24}, showing the average twist and associated
standard deviations as they vary with $A_c$. The mean and median
values of all points are 4.0 and 3.6 turns/AU, respectively. If
the point of the largest standard deviation is excluded, they
become 3.8 and 3.3 turns/AU, respectively. For half of the events of
average twist less than the median value, the standard deviations
are small, indicating a flat profile of $\tau(A)$. Another general
trend is that the larger size the flux rope is as indicated by
larger value of $A_c$, the smaller and  less variable the twist
becomes.

\subsection{Constant-Twist Gold-Hoyle (GH) Flux Rope Model}\label{sec:GH}
The constant-twist or so-called Gold-Hoyle (GH) flux-rope model was originally
developed by \citet{1960MNRAS.GH}. It possesses rather simple and
elegant forms for the magnetic field components in axi-symmetric
cylindrical coordinate ($r,\phi,z$)\citep{1999AIPCF}
\begin{eqnarray}B_z&=&\frac{B_0}{1+T_0^2 r^2}\label{eqGH1}\\
B_\phi&=&\frac{B_0 T_0 r}{1+T_0^2 r^2}.\label{eqGH2}\end{eqnarray}
Here the field-line twist by definition,
$\frac{1}{r}\frac{B_\phi}{B_z}=T_0=2\pi\tau_0$, is strictly
constant and is in the unit of radians/AU, which is also a signed
quantity indicating the chirality of the flux rope. The parameter
$B_0$ corresponds to the axial magnetic field at the center of the
flux rope ($r=0$) which is set to be $B_{z0}$ from the GS results
as given in Table~\ref{tbl:GS}. They usually correspond to the
maximum axial field during the interval (see
Figure~\ref{fig:var_e}). The center of the flux rope is determined
from the GS result as well and since we are only interested in
deriving an approximation of field-line length as function of $A$,
we don't need to explicitly calculate $r$. The length can be
expressed explicitly as a function of $A$ thus can be directly
estimated for each $A$ value obtained from the GS reconstruction.
The other parameter, $\tau_0$, also given in Table~\ref{tbl:GS}
for each event, is obtained by taking the mean value of $\tau$ for
the outer loops where
the twist variation is minimal,  excluding the central core of each flux rope as we discussed earlier in
Section~\ref{sec:GS}. Largely based on the GS reconstruction
results, the GH model provides an alternative and additional means
of estimating, especially extrapolating field-line lengths some of
which are not available through the direct GS model estimates.

%\subsection{Field-line Length Estimates}
 From the GH model, because of the simple forms of the
magnetic field components and the axisymmetric geometry, a flux
function can be derived analytically
\begin{equation}
A(r)=-\frac{B_0}{2T_0}\ln(1+T^2_0r^2).\label{eq:GHA}
\end{equation}
Subsequently, the field line length per AU (i.e., for a section of
the cylinder with an axial length $L_{eff}=1$ AU) can be written
as a function of $A$ ($T_0\equiv 2\pi\tau_0$)
\begin{equation}
L_{GH}=\mathrm{e}^{-T_0A/B_0}=\sqrt{1+T_0^2 r^2}, \label{eq:LGH}
\end{equation}
which tends to increase linearly with radial distance $r$ from the
center of the flux rope when $T_0r\gg 1$. It is also worth noting
that the GH model corresponds to a nonlinear force-free configuration with
the  non-constant force-free parameter
$\alpha=\frac{2T_0}{1+T_0^2 r^2}$, varying with radial distance,
i.e., along $A$ shells as well, as originally derived by
\citet{1960MNRAS.GH}.

Table~\ref{tbl:Le} summarizes the analysis results of measured and
derived magnetic field-line lengths inside the selected MCs
examined by \citet{2011JGRAK}. The entries of Date (1st column),
Type III radio emission times (2nd column), measured field-line
path lengths $L_e$ and D (3rd and 4th columns) are taken from
Table~1 of \citet{2011JGRAK}. The path lengths D were obtained via the inverse-beta approach and were deemed inferior to the measurements $L_e$ by the direct travel-time dispersion analysis.
There are a few unacceptable values of D$<1$ AU. The uncertainty in $L_e$ presented here is owing to the uncertainty in the exact timing of the arrival of
the energetic electrons. The last two columns list the
corresponding estimates of field-line lengths based on the direct
GS reconstruction output, $L_s$, and the GH model approximation,
$L_{GH}$, respectively. The latter is obtained by applying the
equation~(\ref{eq:LGH}) with the necessary parameters supplied by the GS
reconstruction results, i.e., the parameters $\tau_0$ and $B_0=B_{0z}$
from Table~\ref{tbl:GS} and $A=A'$. The corresponding electron
burst onset times at 1 AU are also given in the 2nd column inside
the parentheses. Note that only the event dates and times within
the GS reconstruction intervals as indicated in Table~\ref{tbl:GS}
are listed for which our analysis can yield at least one estimate
of $L_s$ and $L_{GH}$. The others are considered outside of MC
interval hence no analysis results are available. For the times
listed which correspond to locations inside the MC but outside the
loop boundary $A'=A_c$, the estimates of $L_s$ are not available
while the estimates based on the GH model approximation can still be
obtained. We defer detailed comparisons among these length
estimates and discussion of their implications to
Section~\ref{sec:summ}.

\section{Case Studies}\label{sec:CS}
In what follows three events are chosen to be presented as
detailed case studies. The event 1 and 2 are selected because they
possess the maximum number of electron burst onsets inside the MCs
among all the events. The event 7 also contains a modest number of
electron onset times and represents an extreme case of relatively
and persistently long measured path lengths  $L_e$ throughout the
MC interval. Thus these events facilitate a direct and broad
comparison between measured $L_e$ and estimated path lengths based
on the GS reconstruction results and the GH model approximations.

\subsection{Event 1: 18 October 1995}\label{subsec1}
This event was also presented in \citet{1997GeoRLL} and
\citet{2011JGRAK}, which possesses the maximum number of electron
burst occurrences throughout the MC interval. The in-situ
signatures of an MC structure are also strong, as seen from
Figure~\ref{fig1018_ALL}a. The magnetic field magnitude is
elevated and remains around 20 nT, the rotation in the  GSE-Z
component is the largest and clearly seen, and the plasma $\beta$
is fairly low $\sim 0.1$, even after taking into account the
electron temperature contribution ($T_e/T_p\sim 5$). This is a
relatively strong and long-duration MC event with a constant speed
profile and dominant magnetic field, indicating a typical
flux-rope type magnetic structure embedded. This event was also
examined by \citet{2002JGRAHu} as a typical MC event to showcase
the first application of the GS reconstruction method to the
large-scale quasi-static MC flux-rope structures observed in-situ
at 1 AU. The general presentation of the GS results is shown in
Figure~\ref{fig1018_ALL}b and c. The data and a functional fitting
to the transverse pressure $P_t=p+B_z^2/2\mu_0$, the sum of the plasma pressure and the axial magnetic pressure, versus the flux
function $A$ as obtained along the spacecraft path at $y=0$ are
given in Figure~\ref{fig1018_ALL}b, together with a fitting residual $R_f$ indicating the goodness-of-fit of $P_t(A)$ \citep{2004JGRAHu}.
Figure~\ref{fig1018_ALL}c shows the typical presentation of the cross-sectional map of the flux-rope structure as a contour plot of $A(x,y)$ with the axial
component $B_z$ superposed in color. It can also be viewed as a projection of the winding magnetic field lines lying on different iso-surfaces of $A$ ($A$ shells) onto the $(x,y)$
plane. Figure~\ref{fig:3D} shows a rending of the 3D view with a few selected field-lines including the ones rooted on the locations of electron burst onset observations inside
the surface $A'=A_c$.  Therefore the projected field lines that complete multiple turns around the $z$ axis will appear as the closed loops on the cross-sectional map of
Figure~\ref{fig1018_ALL}c enclosed by the outermost loop highlighted in white where $A'\equiv A_c$. There are five incidences of electron burst onsets along $y=0$ as marked by
cross signs with three occurring inside (in black) and the other two outside (in white) of the white loop.

Figure~\ref{fig:lgth1} shows the double-axis plot of the
distributions of field-line twist estimates (left axis) and the corresponding
lengths (right axis) versus the shifted flux function $A-A_0$, including the
available  measurements of $L_e$ with uncertainties, scattered at
different $A$ shells within the flux rope. Note that the shifted
flux function is signed in this plot with the flux-rope center
always located at $A=A_0$. Therefore the sign of the shifted flux
function simply indicates the chirality of the flux rope: negative (positive)
means right (left)-handed. The black thick curve and the
three colored thin curves (see the legend in the top right-hand
corner) represent the field-line twist estimates based on the
graphic method and the other three approximate methods utilizing
the magnetic flux and relative magnetic helicity content
estimates, as described in details by \citet{2014ApJH}. The
graphic method yields the most accurate estimate but is limited to
the inner region of loops satisfying $A'<A_c$. The results
for the other three methods are only for reference purposes to
visually inspect whether they follow the graphic method and the
general trend of the twist distribution beyond the boundary where
the graphic method ceases to provide an estimate \citep{2014ApJH}.
As discussed earlier in \citet{2014ApJH}, the estimate by $-d\Phi_p/d\Phi_t$
(green curve) would exhibit erroneous behavior of rapid rise
toward the boundary of the flux rope (large $A'$ values), as seen
here, due to the rapid decrease in the estimate of $d\Phi_t$, but
not in $d\Phi_p$. Overall, the twist distribution remains fairly
low and constant with larger variations near the flux-rope center,
yielding $\tau_0=1.6$ turns/AU for this case as indicated by the horizontal dashed line.

The corresponding length estimates (magenta curves and dotted
curves) and measurement of $L_e$ (thick black horizontal and
vertical lines) are overplotted versus the shifted flux function
with the scales given by the right axis. There are two sets of
estimates in this case. The thinner ones rise from 1 AU at $A=A_0$
and increase toward the outer loops and they correspond to the
estimates by using the default value $L_{eff}=1$ AU. They do not
intersect the measured $L_e$ at the locations along the $A$ shells
marked by the cross signs. For this particular case, there were
five incidences of electron burst onset occurring within the GS
interval with one occurring very close to the flux-rope center.
There is a short vertical thick black line and a cross at
$A-A_0\approx0$, indicating the range and average of $L_e$ measured at
that location. This measurement at this location enables us to
determine $L_{eff}$ for this flux-rope event since the field line
at the location near the center of the flux rope is mostly
straight. Therefore a direct measurement of $L_{eff}$ is obtained
in this case by taking the mid-point of the range of $L_e$
measured, $L_{eff}\approx1.3$ AU. Then the actual field-line length
estimates of both $L_s$ and $L_{GH}$ are raised from their default
values by simply multiplying the $L_{eff}$ determined, resulting
in the set of thick magenta and dotted curves. These corrected
values will be used in the summary comparison with measurements.
They now intercept a majority of vertical thick black lines except
the one of the largest $L_e$ at a location near the boundary. This adjustment by an
$L_{eff}>1$ AU only applies when such  a direct measurement of $L_e$
is available near the center of a flux rope. We choose the
criterion for such  locations along the $A$ shells satisfying
$|A-A_0|<10$ T$\cdot$m.

\subsection{Event 2: 18 September 1997}\label{subsec2}
This MC event has a very long duration, about two and a half  days as seen
in Figure~\ref{fig0918_ALL}a. The speed is fairly low, around
300-350 km/s during the GS interval. There are significant
variations in the proton temperature $T_p$ (the black trace in the
third panel of Figure~\ref{fig0918_ALL}a) which does not show
clear decrease inside the GS interval compared with the $T_p$
values immediately outside, indicating the possible presence of significant
plasma pressure. This results in a fairly modest plasma $\beta$
$\sim$0.5 within the GS interval, which is still depressed due to
the relatively strong magnetic field magnitude. The plasma
pressure becomes comparable to the axial magnetic pressure near
the middle of the interval as shown in the last panel of
Figure~\ref{fig0918_ALL}a, albeit there is less variation (smaller
gradient) in the plasma pressure. The corresponding $P_t(A)$ plot
and the corresponding cross-sectional map are given in
Figure~\ref{fig0918_ALL}b and c, respectively. Again this is a
right-handed flux rope with a cross-sectional size of about 0.25
AU across. The inner loops enclosed by the  white thick loop
in Figure~\ref{fig0918_ALL}c occupies an area of a diameter about
0.1 AU. In this case, all the electron onset locations are outside
of the closed loops bounded by the white loop where $A'\equiv A_c$
except for one point barely touching this boundary. Therefore most
of the length estimates have to be obtained by the GH model-based
extrapolation.

The corresponding results including the field-line twist
distribution and the actual measurements $L_e$ are shown in
Figure~\ref{fig:lgth2}, in the same format as
Figure~\ref{fig:lgth1}. The twist distribution remains fairly
constant, especially in regions farther away from the flux-rope
center, yielding $\tau_0\approx3.6$ turns/AU. The field-line
length estimates $L_s$ rises from 1 AU at $A'=0$ and increases to
about 1.9 AU at $A'=A_c$, matching the measured $L_e$ at that
location. Beyond that point, no estimates of $L_s$ are available,
but the estimates by $L_{GH}$ are able to continue  as illustrated
by the dashed curve as $A'$
increases toward the outer boundary of the flux rope. These estimates seem to match the additional
measurements of $L_e$ except for the last point (left-most
vertical bar) which is significantly lower than the estimated
value $L_{GH}\approx 3.8$ AU, denoted by the cross sign at top. In this case, since there is no electron burst
onset measurements close to the flux-rope center, the axial length
of the flux rope is unknown and the default value $L_{eff}=1$ AU
is used to obtain the corresponding field-line length estimates from the flux-rope models.
The agreement with the measurements $L_e$ is reasonable. For most
events examined in this study, we have to adopt this approach. Event 1 presented
earlier and event 7, to be presented in the following subsection, are the
only two exceptions.

\subsection{Event 7: 30 August 2004}\label{subsec3}
Event 7 is also a relatively large-scale event with a duration a
little less than 24 hours, resulting in a relatively large-scale
MC flux-rope structure. The in-situ data given in
Figure~\ref{fig0830_ALL}a indicate a typical MC event: clear
enhancement of the magnetic field magnitude and rotation in
direction, low proton temperature and low proton $\beta$ within
the GS interval. Although the magnetic pressure still dominates,
because the ratio $T_e/T_p$ reaches 10 in the GS interval, the
plasma $\beta$ is modest and in the range 0.1-1.0, owing largely
to the contribution by the electron temperature to the total
plasma pressure. The GS reconstruction results including the
contributions of both $T_e$ and $T_p$ are shown in
Figure~\ref{fig0830_ALL}b and c, in the same format as before. The
$P_t(A)$ curve shows a slight bend-over near the end to the right,
which corresponds to the center of the flux rope as represented by
the maximum $A$ value. This behavior indicates a slight decrease
in axial current density thus a weaker transverse field in the
center. The corresponding cross-sectional map in panel (c)
reflects this behavior with the transverse field nearly
vanishing near the center whereas the axial field $B_z$ maintains
a strong and flat distribution inside a large area enclosed by the
inner white loop. Such a configuration indicates that the field
lines near the center are nearly straight with low twist. The 3D
view of field lines is shown in Figure~\ref{fig:3D_7} where
the inner field line, for example, the one of magenta color, is winding
along the axis to large distance, $\sim 1$ AU, before completing
one full turn. The two black lines rooted on two electron onset
locations (two crosses inside the inner white loop on
Figure~\ref{fig0830_ALL}c) near the center show similar behavior to the magenta line. Both the
field-line length estimates $L_s$ and $L_{GH}$ for these two
locations are available while only the estimate $L_{GH}$ for the
other location  outside of the loop $A'=A_c$ is available.
Figure~\ref{fig:lgth7} shows the twist distribution and various
length estimates along the $A$ shells, similar to
Figure~\ref{fig:lgth1}. Here the adjusted length estimates are
also given for an $L_{eff}\approx 3.0 $ AU, based on one measurement of
$L_e$ close to the flux-rope center where $A'<10$ T$\cdot$m. The
average twist is fairly low, yielding $\tau_0\approx 1.2$ turns/AU
for the outer loops, the lowest among all the events. The one
electron burst onset measurement near the center yields a
unusually large axial length of the flux rope, but the accordingly
adjusted length estimates (thick magenta and dotted lines)  show
better agreement with measurements, matching 2 out of 3 values of $L_e$. The one mis-match at the far left is almost outside of the flux-rope
boundary defined by $A=A_b$ beyond which the flux-rope
interpretation based on the GS solution is less reliable. In other
words, the location of this point could be outside of the MC flux
rope and shall be excluded from the field-line length comparison.

\section{Summary and Interpretation of Results}\label{sec:summ}
In this section, we summarize our analysis results presented in
Table~\ref{tbl:Le} and make direct comparison between the measured
path lengths $L_e$ (and D) and the derived ones from the direct GS
model output $L_s$ and the constant-twist GH model estimates
$L_{GH}$. For these handful of events, the path lengths obtained
from the energetic electron burst onset measurements are in the
range of 1 to 4 AU. A few exceptions exist for results
corresponding to D which are less than 1 AU, thus deemed
unacceptable. The apparent limitations and pitfalls of obtaining D
based on energetic electron beams dispersion were discussed in
several works \citep[e.g.,][]{2006ApJK,2011ApJW} but will not be repeated here. We adopt the
results published by \citet{2011JGRAK} and their approach of
weighing more the measurements of $L_e$ as better approximations
of field-line path lengths.

The derived path lengths from the GS together with the GH flux-rope models are within the same range as $L_e$ but
are subject to an uncertainty in the effective length, $L_{eff}$,
the length of a section of the infinite long cylinder that would
correspond well to the flux-rope structure and the intrinsic
characteristic quantities. Therefore the actual field-line length
estimates  are obtained by multiplying the lengths given for a
section of unit axial length (usually 1 AU) by $L_{eff}$ in AU,
whenever such a determination is available, as described in the
case studies of events 1 and 7 presented in Sections~\ref{subsec1} to \ref{subsec3}.
The uncertainty estimates in $L_s$ and $L_{GH}$ are based on
errors propagated from the uncertainties associated with the
measured electron onset times within the GS intervals.

Figure~\ref{fig:Le} shows the ensemble  distribution of measured field-line
path length $L_e$ along the $A$ shells and the one-to-one
comparison between $L_s$ (and $L_{GH}$) and $L_e$ for all events.
Figure~\ref{fig:Le}a shows collectively all the measured $L_e$
along the $A$ shells within GS intervals and their associated
uncertainties. They exhibit a general trend of increasing path
lengths with increasing $A'=|A-A_0|$, i.e., with increasing radial
distance away from the flux-rope center where $A'\equiv 0$. For
the events located near the center (to the left of the vertical
dashed line of $A'=10$ T$\cdot$m) the path
length $L_e$ would represent a direct measurement of $L_{eff}$. For
example, the one closest to $A'=0$ at $L_e\approx 1.3$ AU
corresponds to event 1 presented in Section~\ref{subsec1}. The
ones clustered around $L_e\approx 3.0$ AU correspond to event 7
discussed in Section~\ref{subsec3}. Their locations on the
cross-sectional maps of GS reconstruction results are close to the
center of the flux ropes where the field-line twist values are
small. A few guide lines are also drawn to further elucidate the
trend and the coverage of the constant-twist GH model estimates.
From the GS reconstruction results, an ensemble of field-line
twists is obtained and presented in Table~\ref{tbl:GS} and
Figure~\ref{fig:tau24}, for example, from which a mean value of
twist, as well as the minimum and maximum values are obtained. They are
utilized to provide an estimate of coverage by the area bounded by
the curves based on equation~(\ref{eq:LGH}) varying along $A$
shells for a given constant twist. In Figure~\ref{fig:Le}a, the
set of blue (red) curves corresponds to the length distribution
along $A$ shells based on the GH model for a constant twist of the
minimum, mean, and maximum value from all field-line twist
estimates, respectively, for $L_{eff}=1$ AU (2 AU). In particular, the length
variations for the mean twist values are drawn by dashed lines.
Therefore it can be seen that the majority of the measurements
falls within the region with the lower and upper bound provided by
the GH model of the minimum and the maximum twist and for
$L_{eff}\in [1,2]$ AU. One exception is the measurements from
event 7 as we discussed earlier which might be an indication that
the effective length could reach 3 AU in extreme cases.

Figure~\ref{fig:Le}b shows the direct comparison of $L_s$ versus
$L_e$ with associated uncertainties. Due to the limitation of the
direct field-line length estimate from the GS reconstruction
results, only 9 pairs of data points are available (the 5th column
in Table~\ref{tbl:Le}). It shows good one-to-one correlation,
especially considering that the correlation may be further
improved because the low points of low $L_s$ values beneath the dashed diagonal line
could be raised by a possible correction of
being multiplied by an $L_{eff}> 1$ AU. The same comparison with
expanded length estimates including additional GH model estimates
is shown in Figure~\ref{fig:Le}c where the additional pairs of
$L_{GH}$ and $L_e$ are marked by a cross and in black (the last
column of Table~\ref{tbl:Le}). The correlation deteriorates
compared with panel (b). However the number of outliers from the
one-to-one line is few, especially counting only the ones above
the dashed line, about 3, out of a total of 18, for the reason discussed above regarding panel
(b). For completeness, we also show the same set of results and
comparisons with $L_e$ replaced by D (4th column in
Table~\ref{tbl:Le}) in Figure~\ref{fig:D}. The alternative length
estimates D were provided by \citet{2011JGRAK}  without error
estimates and were not used in their comparison. The agreement of
various model length estimates with the measurements seem to
degrade compared with the previous figure. For instance, the
number of points above the dashed line increases in both panels (b)
and (c).

\section{Conclusions and Discussion}\label{sec:con}
\begin{figure}
 \noindent\includegraphics[]{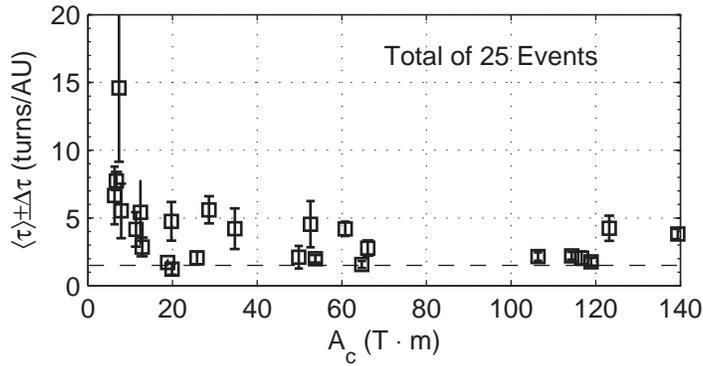}
 \caption{The mean and standard deviation of  field-line twist $\tau(A)$ of 25 magnetic flux-rope events. Each data point with associated error bar is plotted at the corresponding value of $A_c$. }
 \label{fig:tau24}
 \end{figure}

\begin{figure}
\begin{minipage}[b]{.6\textwidth}
\includegraphics[width=1.\textwidth]{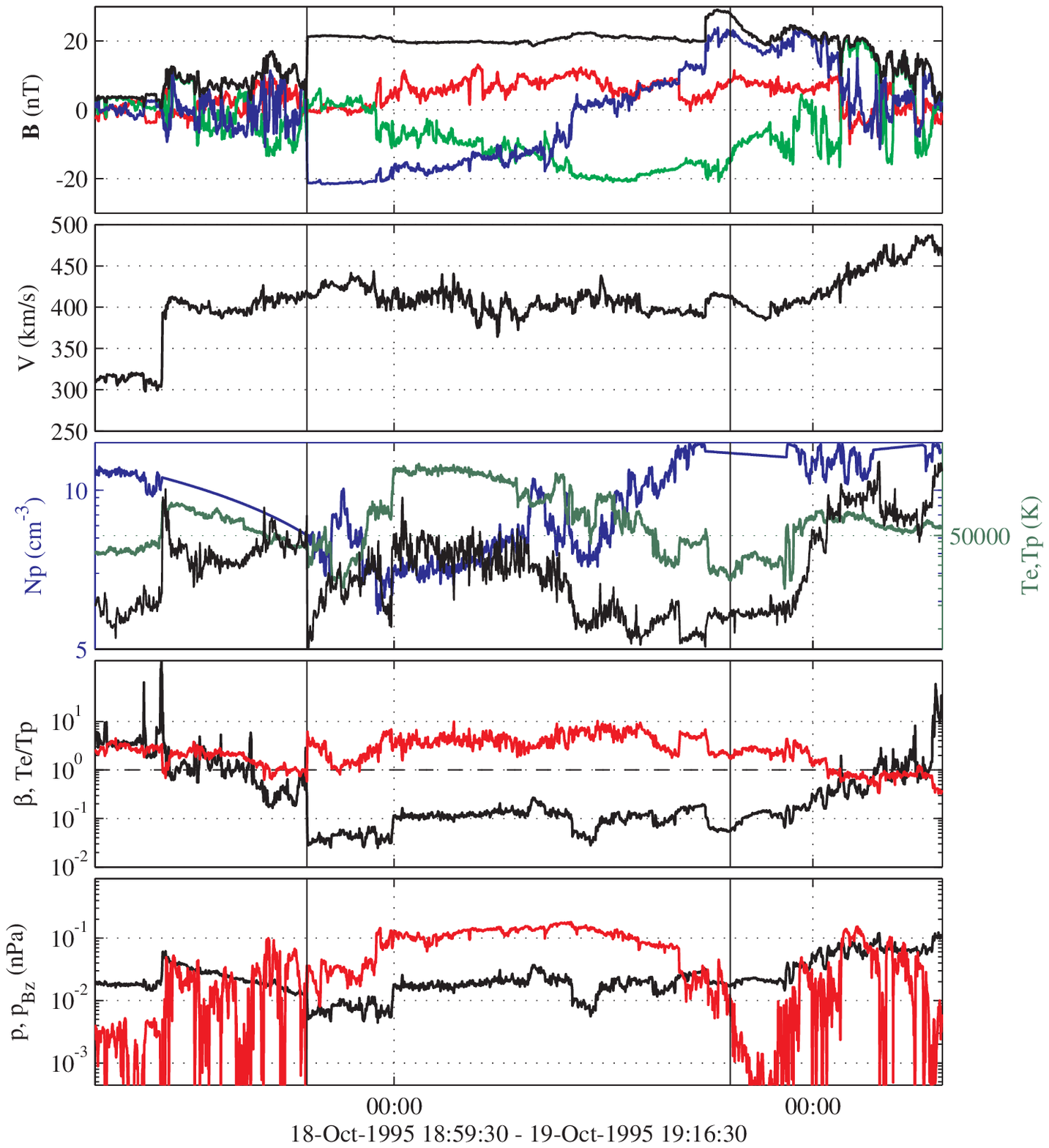}
\centering (a)
\end{minipage}
%\caption{} \label{fig0914_data}
%\end{figure}
%\begin{figure}
\begin{minipage}[b]{.4\textwidth}\centering (b)
\includegraphics[width=.8\textwidth,clip=]{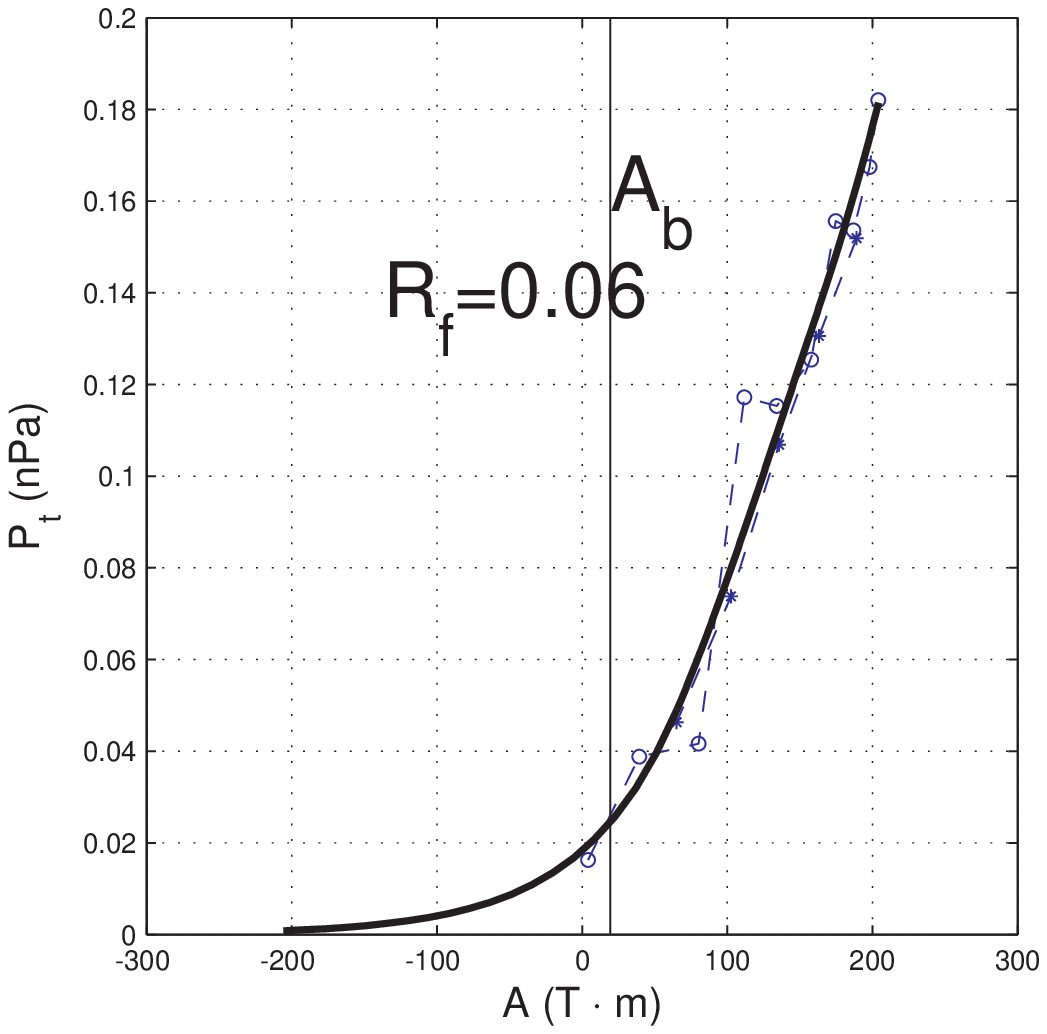}
\includegraphics[width=1.\textwidth]{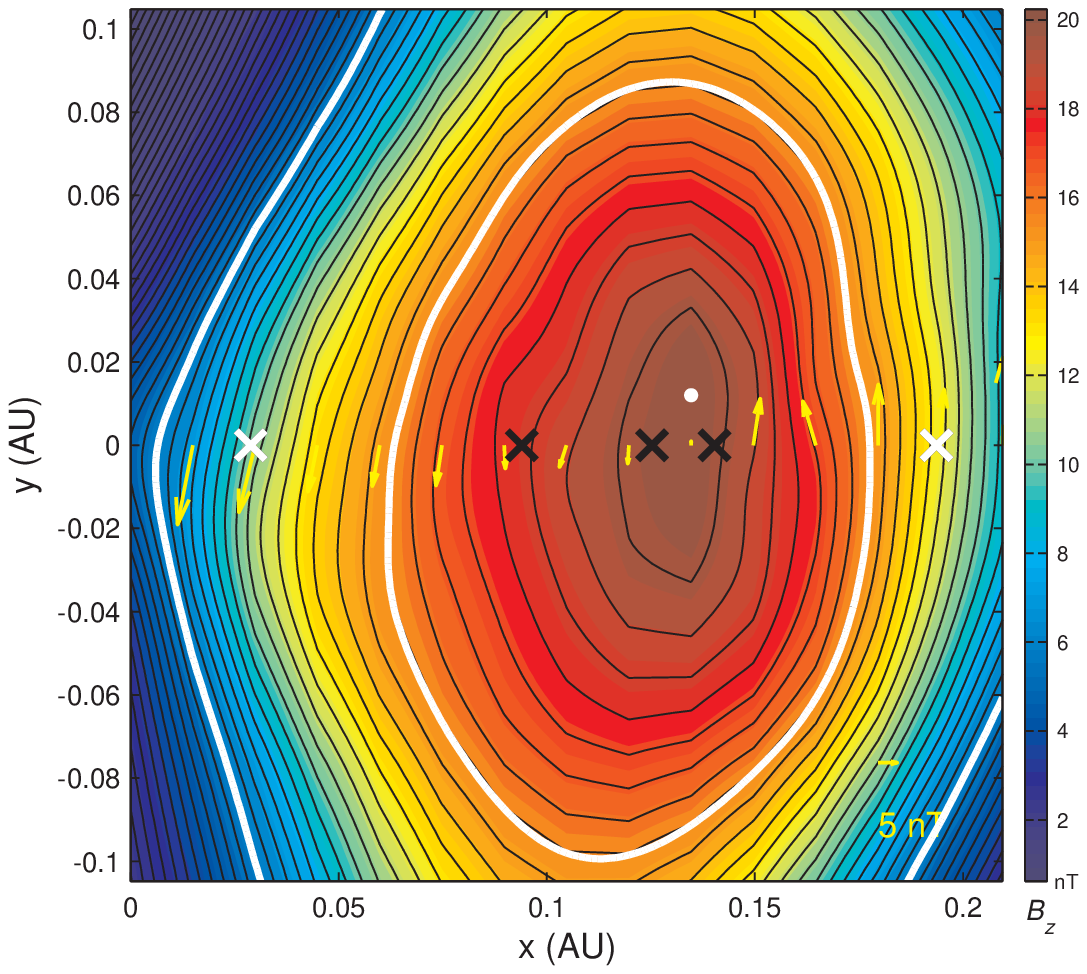}
(c)
\end{minipage}
%\vspace{-1.\textwidth}   % Shift close to the panel top
%     \centerline{%\Large \bf     % Includes the labels (here needs the color package)
%      %\hspace{0.04 \textwidth} \color{white}{(a)}
%      \hspace{0.92\textwidth}  {(b)}
%         \hfill}
%     \vspace{1.\textwidth}
\caption{\linespread{1.0}\small{The GS reconstruction result for event 1 in
Table~\ref{tbl:GS}. (a) Time series of Wind spacecraft
measurements: (from top to bottom panels) the in-situ magnetic
field magnitude (black) and GSE-X (red), Y (green), and Z (blue)
components, the plasma bulk flow speed, the proton density (left
axis; blue) and proton (black) and electron (green; if available) temperature
(right axis), the plasma $\beta$ (black) and the electron over proton
temperature ratio (red; if available), and the plasma and axial
magnetic field (red) pressure.  The vertical lines mark the GS
reconstruction interval as given beneath the last panel. (b) The
measurements of $P_t(x,0)$ versus $A(x,0)$ and the fitted $P_t(A)$
curve (thick black line). The flux rope boundary is marked at
$A=A_b$ and a fitting residue $R_f$ is denoted. (c) The
cross-sectional map of the solution $A(x,y)$ (black contour lines)
and the axial field $B_z(A)$ (filled  contours in color). The
yellow arrows are the measured transverse magnetic field along the
spacecraft path ($y=0$). The white contour lines denote the
boundary $A=A_b$ (outer) and $A'=A_c$ (inner) while the white dot
denotes the center where the axial field is the maximum and
$A\equiv A_0$. The crosses along $y=$ denote the locations where
the electron burst onsets were observed. The ones inside the  white loop are in black.}} \label{fig1018_ALL}
\end{figure}

\begin{figure}
 \noindent\includegraphics[width=8.3cm]{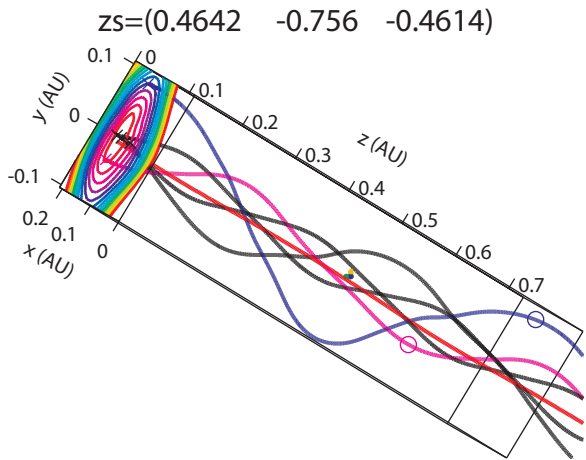}
 \caption{3D view of the flux-rope structure toward Sun for Event 1 with selected field lines. The view angle is such that north is upward and the ecliptic plane is horizontal.
 Black lines are the field lines rooted at the footpoints where the electron onsets were observed. Circles mark the locations where the field lines complete one full turn around the $z$ axis.
The orientation of $z$ axis is given on top in GSE coordinate.}
 \label{fig:3D}
 \end{figure}

\begin{figure}
 \noindent\includegraphics[width=8.3cm]{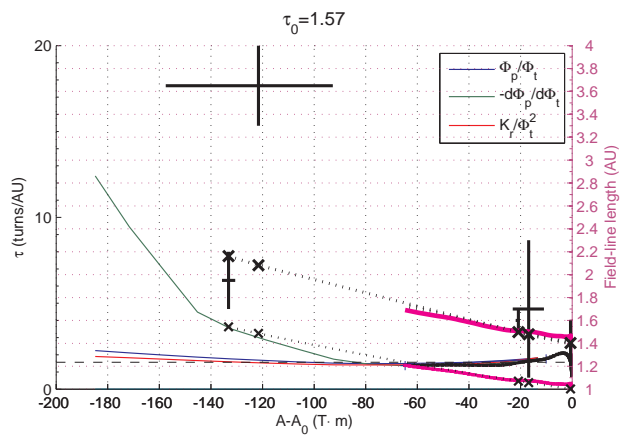}
 \caption{Field-line twist  and length  distributions along the $A$ shells for Event 1. The twist values are indicated by the left axis with the estimate by the graphic method
 in thick black curve. The results from the other three approximate methods are given by the thin blue, green and red curves, respectively, as indicated by the legend. Each curve ends at
 $A=A_b$.
 The horizontal dashed line denotes $\tau_0$ as given on top of the plot in turns/AU. The magenta lines represent the length estimate $L_s$, the dotted lines $L_{GH}$,
 and the black thick verticle lines $L_e$ with uncertainties. All the length scales are given by the right axis. The crosses mark the locations along the $A$ shells where the measurements
 of $L_e$ were obtained, and the corresponding estimates of $L_{GH}$. The set of thinner magenta and dotted curves originating from 1 at $A'=0$ corresponds to the default value
 $L_{eff}=1$ AU. The other set corresponds to $L_{eff}=1.3$ AU. }
 \label{fig:lgth1}
 \end{figure}

\begin{figure}
\begin{minipage}[b]{.6\textwidth}
\includegraphics[width=1.\textwidth]{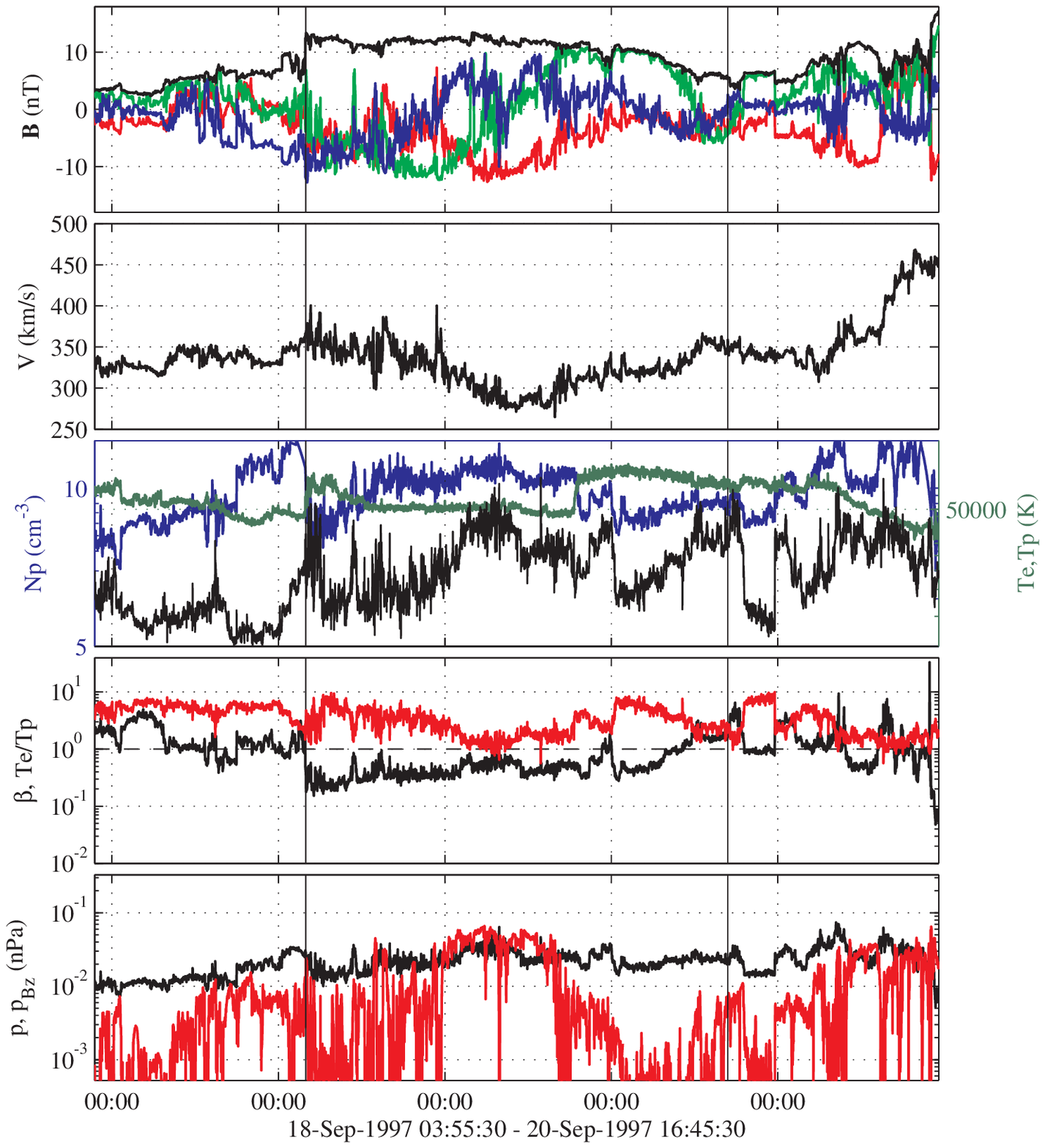}
\centering (a)
\end{minipage}
%\caption{} \label{fig0914_data}
%\end{figure}
%\begin{figure}
\begin{minipage}[b]{.4\textwidth}\centering (b)
\includegraphics[width=.8\textwidth,clip=]{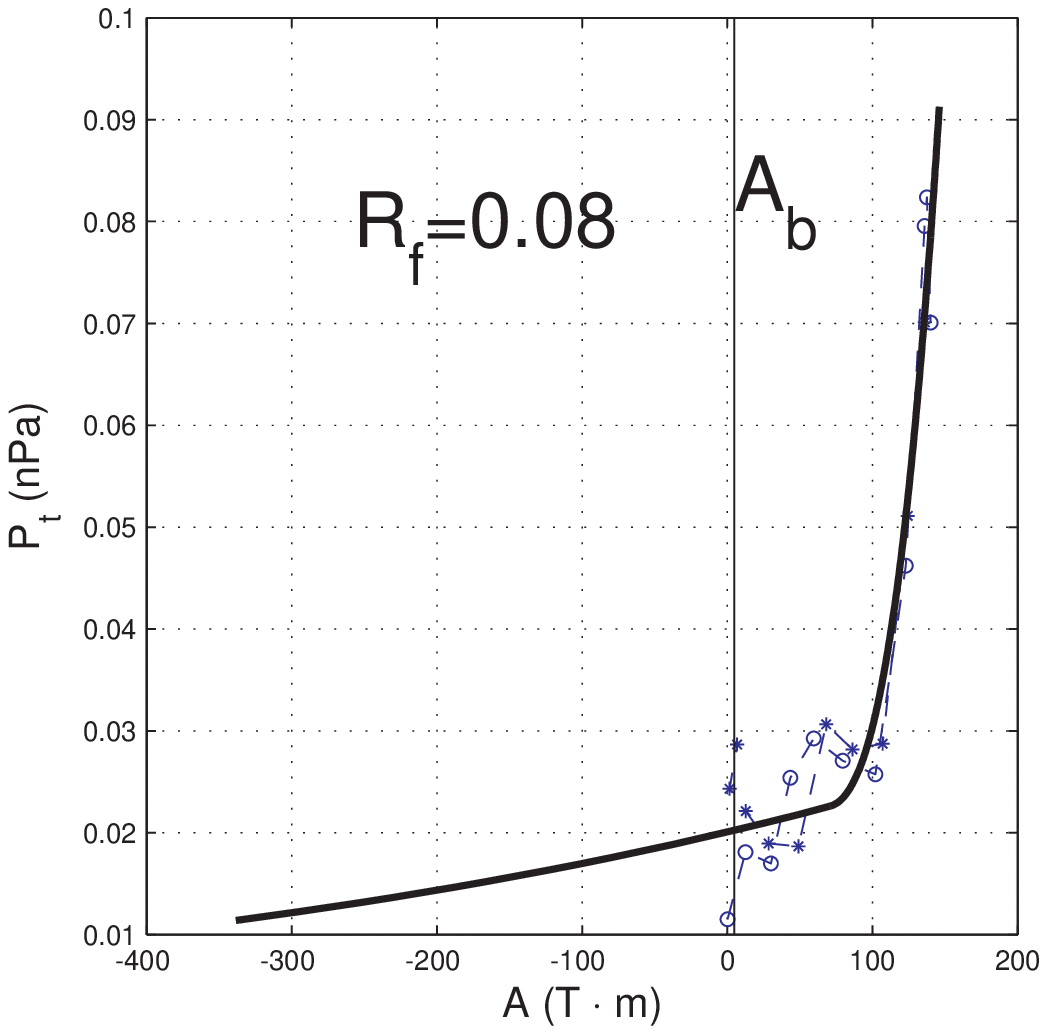}
\includegraphics[width=1.2\textwidth]{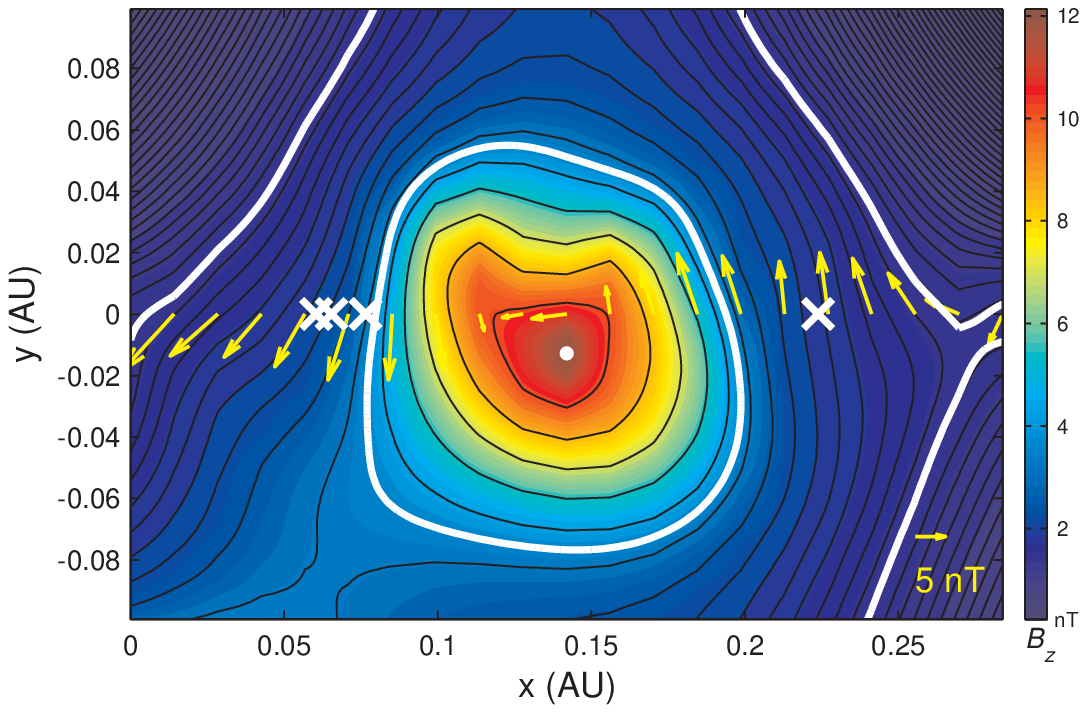}
(c)
\end{minipage}
%\vspace{-1.\textwidth}   % Shift close to the panel top
%     \centerline{%\Large \bf     % Includes the labels (here needs the color package)
%      %\hspace{0.04 \textwidth} \color{white}{(a)}
%      \hspace{0.92\textwidth}  {(b)}
%         \hfill}
%     \vspace{1.\textwidth}
\caption{The GS reconstruction result for event 2 in
Table~\ref{tbl:GS}. Format is the same as
Figure~\ref{fig1018_ALL}.} \label{fig0918_ALL}
\end{figure}

\begin{figure}
 \noindent\includegraphics[width=8.3cm]{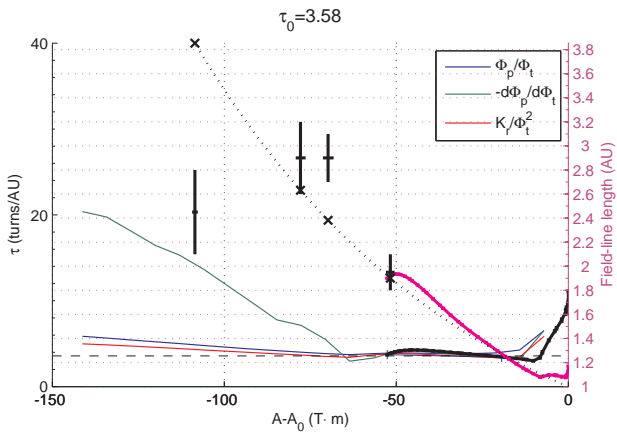}
 \caption{Field-line twist and length distributions along the $A$ shells for Event 2. Format is the same as Figure~\ref{fig:lgth1}.
 Here the effective axial length takes the default value $L_{eff}=1$ AU, only.}
 \label{fig:lgth2}
 \end{figure}

\begin{figure}
\begin{minipage}[b]{.6\textwidth}
\includegraphics[width=1.\textwidth]{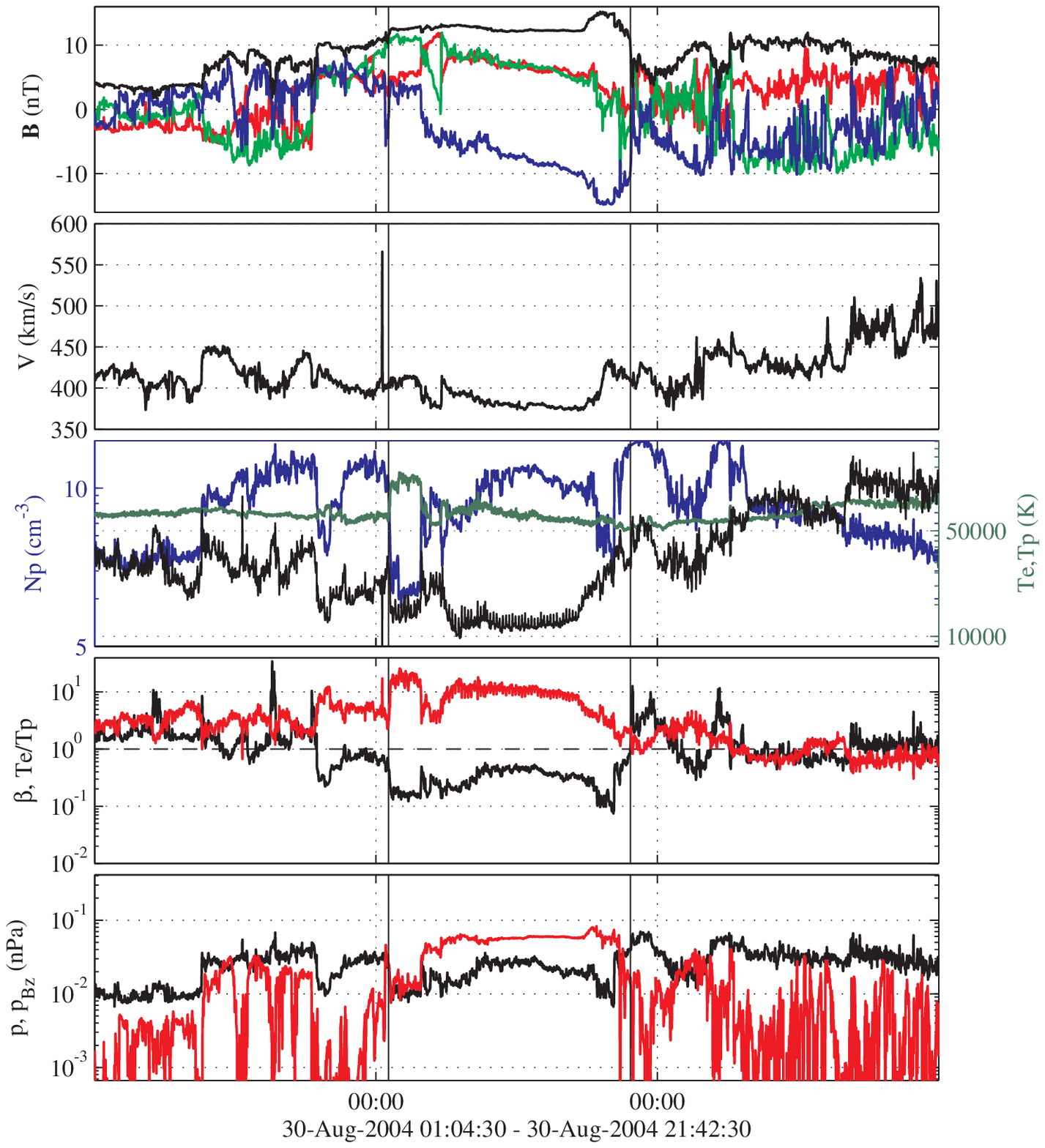}
\centering (a)
\end{minipage}
%\caption{} \label{fig0914_data}
%\end{figure}
%\begin{figure}
\begin{minipage}[b]{.4\textwidth}\centering (b)
\includegraphics[width=.8\textwidth,clip=]{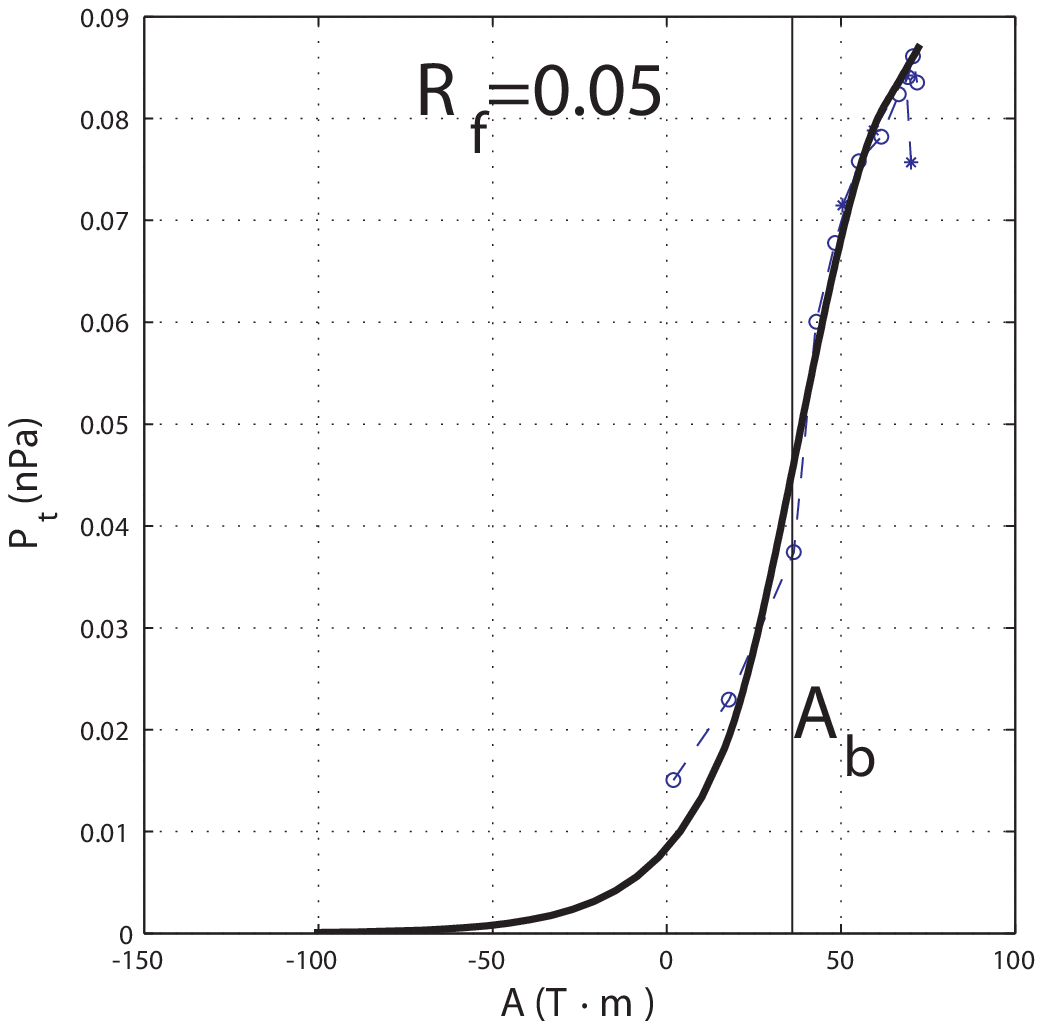}
\includegraphics[width=1.\textwidth]{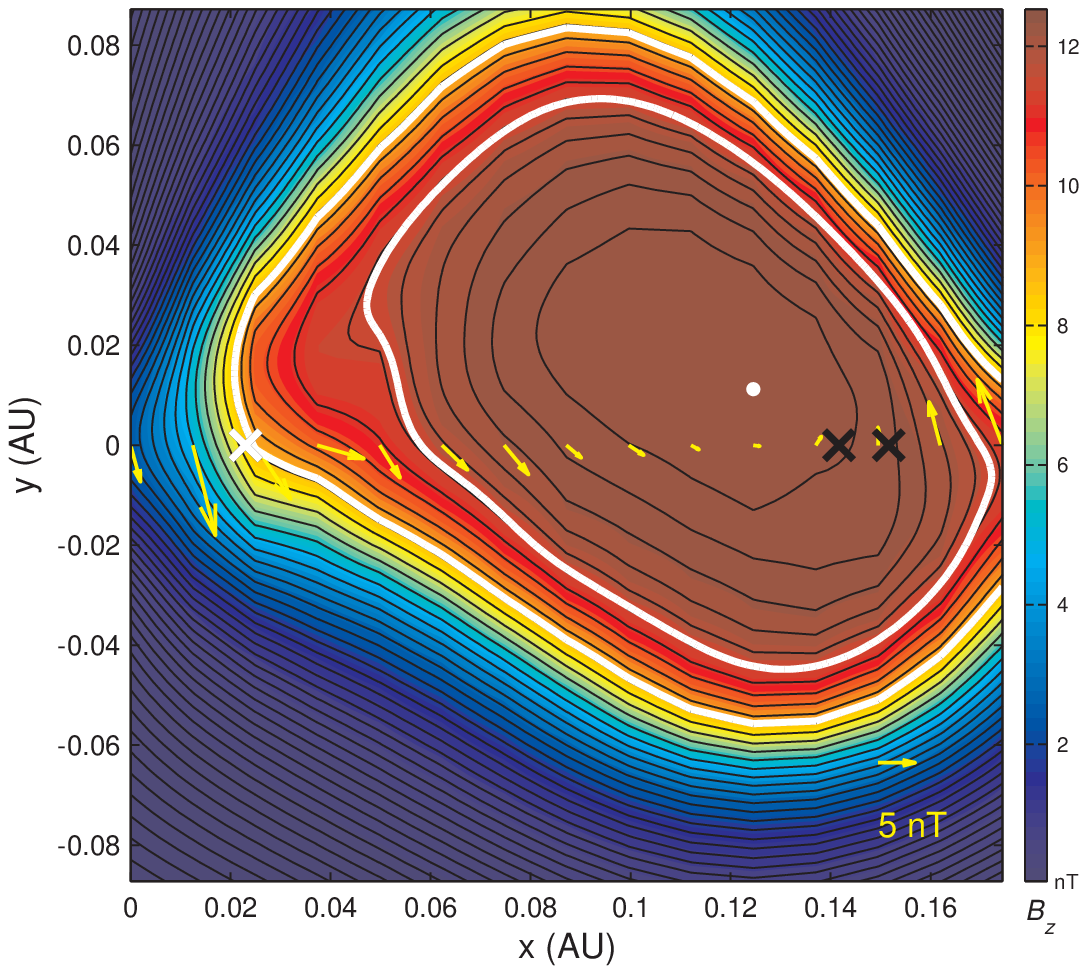}
(c)
\end{minipage}
%\vspace{-1.\textwidth}   % Shift close to the panel top
%     \centerline{%\Large \bf     % Includes the labels (here needs the color package)
%      %\hspace{0.04 \textwidth} \color{white}{(a)}
%      \hspace{0.92\textwidth}  {(b)}
%         \hfill}
%     \vspace{1.\textwidth}
\caption{The GS reconstruction result for event 7 in
Table~\ref{tbl:GS}. Format is the same as
Figure~\ref{fig1018_ALL}.} \label{fig0830_ALL}
\end{figure}

\begin{figure}
 \noindent\includegraphics[width=8.3cm]{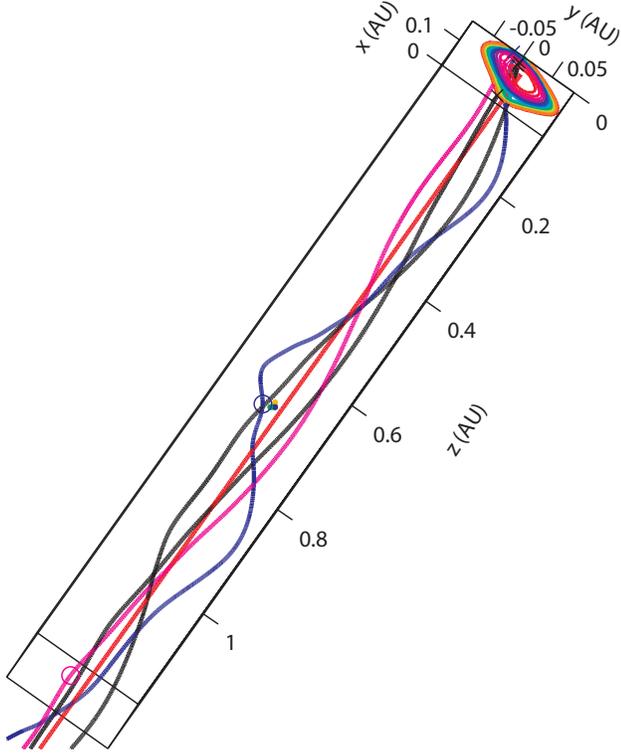}
 \caption{3D view of the flux-rope structure toward Sun for Event 7. Black lines are the field lines rooted at the footpoints where the electron onsets were observed. Format is the same as Figure~\ref{fig:3D}.}
 \label{fig:3D_7}
 \end{figure}

\begin{figure}
 \noindent\includegraphics[width=8.3cm]{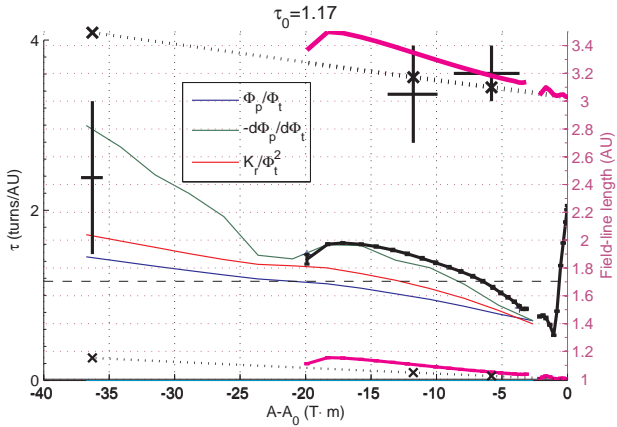}
 \caption{Field-line twist and length distributions along the $A$ shells for Event 7. Format is the same as Figure~\ref{fig:lgth1}. Here the
 set of thicker magenta and dotted curves corresponds to $L_{eff}\approx 3.0$ AU. }
 \label{fig:lgth7}
 \end{figure}

\begin{figure}
 \noindent\includegraphics[width=.8\textwidth]{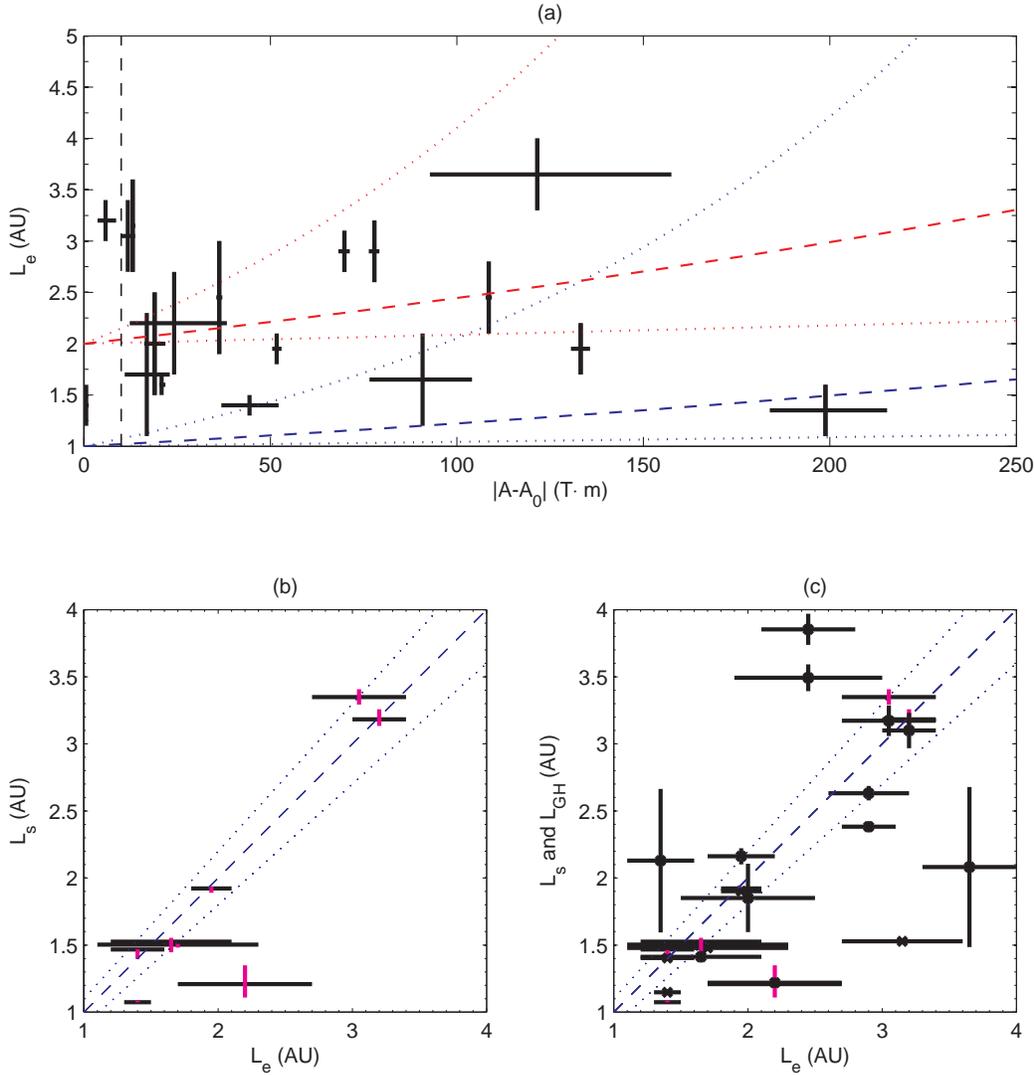}
 \caption{Summary and comparison of field-line length estimates with $L_e$: (a) the ensemble of measurements $L_e$ vs. $A'$ for all events; the dashed and dotted lines mark the variations of
 field-line lengths of GH model with certain constant twist, and the vertical dashed line denotes $A'=10$ T$\cdot$m (see text for details), (b) the one-to-one plot of
 $L_s$ (in magenta) vs.
 $L_e$, and (c) the one-to-one plot of both $L_s$ and $L_{GH}$ vs. $L_e$ where the latter is marked by  black cross signs and  vertical lines. In panels (b) and (c), the dashed
 line denotes the one-to-one diagonal line while the dotted lines mark a 10\% uncertainty bound.   }
 \label{fig:Le}
 \end{figure}

\begin{figure}
 \noindent\includegraphics[width=.8\textwidth]{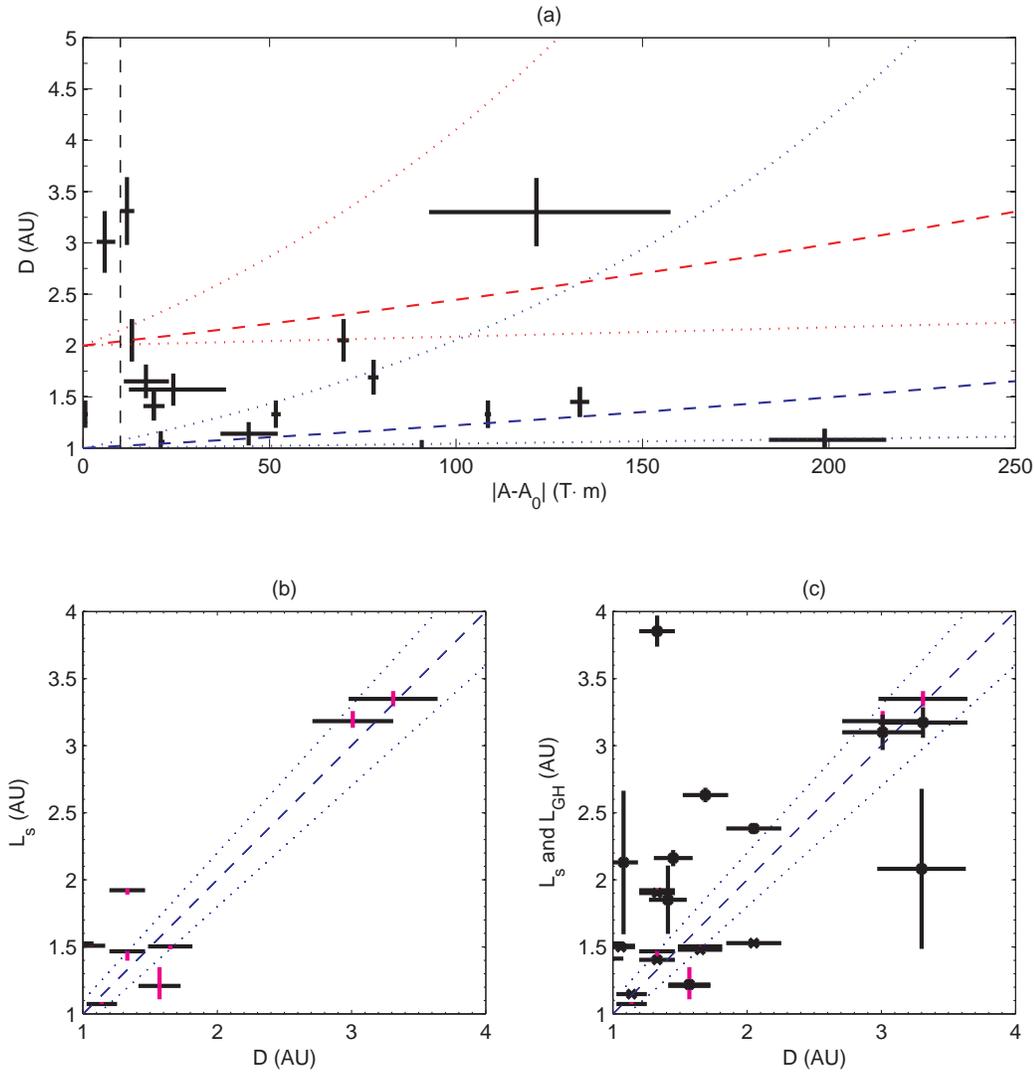}
 \caption{Summary and comparison of field-line length estimates with D which were given by
 \citet{2011JGRAK} without associated uncertainties. Here a
 uniform 10\% uncertainty in D is assumed. Format is the same as
 Figure~\ref{fig:Le}.}
 \label{fig:D}
 \end{figure}

In conclusion, we have examined the flux-rope structures embedded
within 7 MC events, in particular the field-line length and twist
distributions, based on the GS reconstruction method and the
constant-twist GH flux-rope model. We carry out direct comparison
of field-line length estimates with the unique measurements of
field-line path lengths  obtained from timing observations of
energetic electrons traveling along individual field lines from Sun to Earth. We limit
our analysis to the same set of MC events reported by
\citet{2011JGRAK} and employ their published measurements of $L_e$ to
facilitate a highly comparative study but with different flux-rope
models. Our conclusion, somewhat in contrary to Kahler's, is that
the flux-rope interpretation of the magnetic structures embedded
within MCs is largely consistent with the analysis of direct
comparison between the modeled field-line length estimates and the
direct measurements $L_e$. The correlations between $L_e$ and
$L_s$ (and $L_{GH}$) are well established as seen in
Figure~\ref{fig:Le} and the field-line length does exhibit a
general trend of increasing from the flux-rope center. Such a
trend as displayed in Figure~\ref{fig:Le}a as a function of the
shifted flux function has general implication for a flux-rope
structure independent of specific models. On the other hand, we
agree with \citet{2011JGRAK} and others on that such a comparison
provides additional evidence for the inconsistency of Lundquist
model in characterizing the flux-rope structures observed in-situ
at 1 AU. As we indicated in \citet{2014ApJH}, the magnetic field-line twist distribution
within MC flux ropes often exhibits inconsistency with the
Lundquist model, but better supports the GH model of a constant
twist (see Figure~\ref{fig:tau24} and associated descriptions).
The present study further supports the findings of such inconsistency and
provides additional support for the GH model by the direct
comparison of field-line length estimates with the corresponding
measurements. It is also important to show that some electron burst
onset observations are able to provide a direct measurement of the
axial length of the section of a cylindrical flux rope, a critical parameter for the existing flux-rope models. Based on our analysis of a limited number
of events, we
argue that under most circumstances, such a  constraint on the effective axial
length of a cylindrical flux rope is $L_{eff}\in[1, 2]$ AU, which
has significant applications for the relevant studies of deriving and
relating various physical quantities to their solar sources.

It might not be hard to perceive why the comparison with the
Lundquist model failed. Based on the Lundquist model, the
field-line length would increase to infinity at the boundary at
which the axial field vanishes by definition. Therefore the
Lundquist model would yield large path lengths toward the outer
loops of a flux rope. On the other hand, the GH model length would
possess a more modest rate of increase from the center to the
outer boundary of the flux rope, approximately linearly with $r$
as indicated by equation~(\ref{eq:LGH}), remaining finite.
Therefore the correlation between $L_e$ and $L_{GH}$ is more
favorable. As we discussed in \citet{2014ApJH}, the underlying
theoretical consideration for advocating the GH model is that it
describes space plasmas in non-linear force-free state which is
well preserved from its origination from the Sun, propagation
through the interplanetary space to reaching Earth. The ideal
magnetohydrodynamic (MHD) conditions are probably well satisfied
during the processes in the space plasma on the Sun and in the
interplanetary space. The flux surfaces embedded within these
structures remain distinct and well preserved upon their
generation and are not destroyed by finite and highly localized
resistivity, resulting in a non-linear  force-free state as
observed in-situ at 1 AU.

As an ongoing effort, we are extending the analysis to more events
and utilizing more comprehensive sets of available observations.
Some issues not addressed in the present work will be pursued in
the forthcoming studies. For example, generally we would expect
difficulty when the measured path lengths are exceedingly long and
near the flux-rope center as we explained in the case study of
event 7. Our interpretation of a flux-rope structure with an
unusually long axial length of $\sim$3 AU  needs to be further
validated by additional event studies. Another related issue is
what effect there is regarding the finite plasma pressure gradient. A
slight change in the model output of the configuration of the flux
rope would affect the spatial locations of the electron onsets
where the measurements of $L_e$ were taken. Such a change in
location would yield change in the length estimates by specific
models. The amount of change may depend on whether or not the
plasma pressure gradient is taken into account. Therefore a
detailed assessment of the differences as resulted from the GS
reconstruction results (non-force free in general) and the GH
model (nonlinear force-free)  estimates is planned for future work.

%%% End of body of article:

%%%%%%%%%%%%%%%%%%%%%%%%%%%%%%%%
%% Optional Appendix goes here
%
% \appendix resets counters and redefines section heads
% but doesn't print anything.
% After typing \appendix
%
%\section{Here Is Appendix Title}
% will show
% Appendix A: Here Is Appendix Title
%
%%%%%%%%%%%%%%%%%%%%%%%%%%%%%%%%%%%%%%%%%%%%%%%%%%%%%%%%%%%%%%%%
%
% Optional Glossary or Notation section, goes here
%
%%%%%%%%%%%%%%
% Glossary is only allowed in Reviews of Geophysics
% \section*{Glossary}
% \paragraph{Term}
% Term Definition here
%
%%%%%%%%%%%%%%
% Notation -- End each entry with a period.
% \begin{notation}
% Term & definition.\\
% Second term & second definition.\\
% \end{notation}
%%%%%%%%%%%%%%%%%%%%%%%%%%%%%%%%%%%%%%%%%%%%%%%%%%%%%%%%%%%%%%%%
%
%  ACKNOWLEDGMENTS

\begin{acknowledgments}
The work of JQ is supported by NASA Guest Investigator Program
NNX12AH50G. QH  acknowledges NASA grant NNX12AH50G, NRL contract
N00173-14-1-G006, and NSF grant AGS-1062050 for support.
 The Wind spacecraft data are provided by the NASA
CDAWeb (\tt{http://cdaweb.gsfc.nasa.gov/}).

\end{acknowledgments}

%% ------------------------------------------------------------------------ %%
%%  REFERENCE LIST AND TEXT CITATIONS
%
% Either type in your references using
% \begin{thebibliography}{}
% \bibitem{}
% Text
% \end{thebibliography}
%
% Or,
%
% If you use BiBTeX for your references, please use the agufull08.bst file (available at % ftp://ftp.agu.org/journals/latex/journals/Manuscript-Preparation/) to produce your .bbl
% file and copy the contents into your paper here.
%
% Follow these steps:
% 1. Run LaTeX on your LaTeX file.
%
% 2. Make sure the bibliography style appears as \bibliographystyle{agufull08}. Run BiBTeX on your LaTeX
% file.
%
% 3. Open the new .bbl file containing the reference list and
%   copy all the contents into your LaTeX file here.
%
% 4. Comment out the old \bibliographystyle and \bibliography commands.
%
% 5. Run LaTeX on your new file before submitting.
%
% AGU does not want a .bib or a .bbl file. Please copy in the contents of your .bbl file here.

\bibliographystyle{agufull08}
\bibliography{ref_master}

%Reference citation examples:

%...as shown by \textit{Kilby} [2008].
%...as shown by {\textit  {Lewin}} [1976], {\textit  {Carson}} [1986], {\textit  {Bartholdy and Billi}} [2002], and {\textit  {Rinaldi}} [2003].
%...has been shown [\textit{Kilby et al.}, 2008].
%...has been shown [{\textit  {Lewin}}, 1976; {\textit  {Carson}}, 1986; {\textit  {Bartholdy and Billi}}, 2002; {\textit  {Rinaldi}}, 2003].
%...has been shown [e.g., {\textit  {Lewin}}, 1976; {\textit  {Carson}}, 1986; {\textit  {Bartholdy and Billi}}, 2002; {\textit  {Rinaldi}}, 2003].

%...as shown by \citet{jskilby}.
%...as shown by \citet{lewin76}, \citet{carson86}, \citet{bartoldy02}, and \citet{rinaldi03}.
%...has been shown \citep{jskilbye}.
%...has been shown \citep{lewin76,carson86,bartoldy02,rinaldi03}.
%...has been shown \citep [e.g.,][]{lewin76,carson86,bartoldy02,rinaldi03}.
%
% Please use ONLY \citet and \citep for reference citations.
% DO NOT use other cite commands (e.g., \cite, \citeyear, \nocite, \citealp, etc.).

%% ------------------------------------------------------------------------ %%
%
%  END ARTICLE
%
%% ------------------------------------------------------------------------ %%
\end{article}
\newpage
\begin{table}[p]
\caption{GS Reconstruction Results of Selected Wind MC Events from
\citet{2011JGRAK}}\label{tbl:GS} \centering
\begin{tabular}{lccccccccc}
\hline Event \# : GS  Interval  & $\langle\tau\rangle\pm\Delta\tau$&$B_{z0}$ &$\Phi_{t,\mathrm{max}}$ &$\Phi_{p,\mathrm{max}}$ &$K_{r,\mathrm{max}}$ &$\bar\tau_H$&$\bar\tau_F$&$\tau_0$\\
 MM/DD/YYYY hh:mm:ss &    turns/AU & nT & $10^{21}$ Mx& $10^{21}$ Mx& $10^{42}$ Mx$^2$ &  & &\\
 \hline
1: 10/18/1995 18:59:30-(+1) 19:16:30  & $1.57\pm0.26$&
20   &    1.2    &    2.8   &    2.9   &    1.9& 2.2&1.6\\
2: 9/18/1997 03:55:30-(+2) 16:45:30&  $4.55\pm1.71$& 12 &0.36&
2.1   &   0.64   &    5.0 & 5.9&3.6\\
3: 11/6/2000 23:08:30-(+1) 18:46:30 & $2.16\pm0.31$& 25&
0.90   &    2.0   &    1.9  &    2.3& 2.2&2.2\\
4: 7/10/2001 18:31:30-(+2) 07:47:30 & $6.66\pm2.13$& 7.8 &    0.039 &     0.35 &     0.012 &      8.3 &9.0&6.0\\
5: 10/1/2002 00:49:30-19:31:30 & $4.20\pm0.54$& 26 &     0.22 &     0.99 &     0.23 &      4.7 &4.4&5.0\\
6: 7/24/2004 11:56:30-(+1) 07:01:30 & $2.11\pm     0.83$& 22  &     1.6  &     2.9  &     3.5   &    1.4  &     1.8&1.6\\
7: 8/30/2004 01:4:30-21:42:30 & $1.24\pm0.41$&12 & 0.38& 0.55&
0.24&   1.7&    1.4& 1.2\\
 \hline
\end{tabular}
%\tablenotetext{a}{The Date, Type III times, $L_e$ and D are taken
%from Table 1, .}
\end{table}

\begin{table}[p]
\caption{ Measured and Derived Magnetic Field-line Lengths inside
Magnetic Flux Ropes\tablenotemark{a}}\label{tbl:Le} \centering
\begin{tabular}{cccccc}
\hline
 Date  & Type III (e-), UT &$L_e$, AU&D, AU &$L_s$, AU& $L_{GH}$, AU\\
\hline 18-Oct-1995& 19:56 (22:20:35)  &      3.3-4.0&
3.30&       ... &     $2.08\pm0.60$\\
19-Oct-1995& 05:18 (05:53:17)&   1.5-1.7 & 1.06&
1.51-1.51 &      $1.50\pm0.003$ \\
19-Oct-1995& 08:46 (09:30:58)&      1.2-1.6 & 1.33&
     1.40-1.47 & $1.40\pm0.002$\\
 19-Oct-1995& 10:28 (11:15:20)&  1.1-2.3 & 1.65&1.48-1.51 &   $1.48\pm
0.02$ \\
19-Oct-1995 & 16:57 (17:25:50)&    1.7-2.2& 1.45& ...&     $2.16\pm0.06$\\
 18-Sep-1997& 16:06 (16:52:37)&  2.6-3.2&      1.69   &...& $2.63\pm0.05$ \\
18-Sep-1997& 17:09 (17:56:00)&  2.7-3.1& 2.05 &...& $2.38\pm0.04$ \\
18-Sep-1997& 19:51 (20:19:57)& 1.8-2.1& 1.33  & 1.89-1.94&
$1.90\pm0.02$ \\
20-Sep-1997& 03:16 (03:53:15) & 2.1-2.8& 1.33 &...&
$3.85\pm0.12$\\
07-Nov-2000& 00:08 (00:56:14)&  1.1-1.6& 1.08 &...& $2.13\pm
0.54$\\
07-Nov-2000& 15:40 (16:33:22)& 1.2-2.1& 0.98& 1.45-1.55& $1.41\pm0.04$ \\
10-Jul-2001& 22:53 (+1 00:44:47)&  1.5-2.5& 1.41 &...&$1.85\pm0.26$\\
12-Jul-2001&01:11 (02:28:27)&  2.7-3.6& 2.05 &...& $1.53\pm0.02$ \\
%01-Oct-2002& 06:07:10 0.531 -27.6 1.43 1.56 1.43 1.25 1.23 1.19 1.25 0.0195 \\
01-Oct-2002& 09:12 (11:55:21)&  1.7-2.7& 1.57& 1.11-1.35& $1.22\pm0.04$ \\
%01-Oct-2002 15:13:20 0.725 -47 1.61 1.62 1.62 1.43 1.54 1.66 1.47 0.045 \\
24-Jul-2004 & 18:43 (19:16:51) & 1.3-1.5 & 1.14 & 1.07-1.09
&$1.15\pm0.01$\\
30-Aug-2004 & 03:09 (03:44:47)   &    1.9-3.0& 0.54 &...&
$3.49\pm0.10$\\
30-Aug-2004 & 16:13 (17:47:05) & 3.0-3.4& 3.01&3.13-3.26&
$3.10\pm
0.13$ \\
30-Aug-2004& 18:09 (18:57:40)& 2.7-3.4& 3.31& 3.29-3.41& $3.17\pm
0.12$\\
 \hline
\end{tabular}
\tablenotetext{a}{The Date, Type III times, $L_e$ and D are taken
from Table 1, \citet{2011JGRAK}.}
\end{table}

%\begin{figure}
% \noindent\includegraphics[width=8.3cm]{win_3Dwin259g97.eps}
% \caption{3D view of the flux-rope structure toward Sun for Event \#2. The locations of observed electron onsets were all outside of the inner boundary where $A=A_c$.}
% \label{fig:3D}
% \end{figure}
%

%\begin{figure}
% \noindent\includegraphics[width=.5\textwidth]{lgth_win311g00.eps}
% \noindent\includegraphics[width=.5\textwidth]{lgth_win189g01.eps}\\
% \noindent\includegraphics[width=.5\textwidth]{lgth_win271g02.eps}
% \noindent\includegraphics[width=.5\textwidth]{lgth_win205g04.eps}\\
% \caption{(to be deleted) Field-line twist and length distributions along the $A$ shells for Events \#3-6.}
% \label{fig:lgth3-6}
% \end{figure}
%

% See below for how to make sideways figures or tables.

\end{document}